%% file: ms.tex
\documentclass[12pt,preprint]{aastex}

\usepackage{xspace,times,perpage}
\usepackage[usenames]{color}
\usepackage{hyperref}
\usepackage{lscape}
\usepackage{ifthen}
\MakePerPage{footnote}
\begin{document}

\input{01_intro}
\input{02_data}
\input{03_result}

\input{04_conclusion}

\input{05_bib}
\input{table_a}
\input{table_b}
\input{table_c}
\input{figures}

\end{document}

%% file: 01_intro.tex
\title{\bf Black-Body Stars}
\author{Nao Suzuki}
\affil{Kavli Institute for the Physics and Mathematics of the Universe, University of Tokyo,\\
 Kashiwa 277-8583 Japan}
\author{Masataka Fukugita}
\affil{Kavli Institute for the Physics and Mathematics of the Universe, University of Tokyo,\\
 Kashiwa 277-8583 Japan}
\affil{Institute for Advanced Study, Princeton NJ08540, U.S.A.} 

\begin{abstract}
We report the discovery of stars that show spectra very close to the
black-body radiation. We found 17 such stars out of 798,593 stars in
the Sloan Digital Sky Survey (SDSS) spectroscopic data archives.
We discuss the value of these stars for the calibration of
photometry, whatever is the physical nature of these stars.  This
gives us a chance to examine the accuracy of the zero point of SDSS
photometry across various passbands: we conclude that the zero point
of SDSS photometric system is internally consistent across its five
passbands to the level below 0.01 mag.  We may also examine the
consistency of the zero-points between UV photometry of Galaxy
Evolution Explorer and SDSS, and IR photometry of Wide-field Infrared
Survey Explorer against SDSS.  These stars can be used as not only
photometric but spetrophotometric standard stars.  We suggest that
these stars showing the featureless black-body like spectrum of the
effective temperature of $10000\pm1500$K are consistent with DB white
dwarfs with the temperature too low to develop helium absorption features.
\end{abstract}

\section{Introduction}
We report the discovery of stars that have nearly a perfect black-body
spectrum without any features in the wavelength range from ultraviolet
to infrared.  A search is made for objects with featureless spectrum
in the spectral archives of the Sloan Digital Sky Survey
(SDSS)\citep{DR7,DR12}, supplemented with Galaxy Evolution Explorer
(GALEX)\citep{galex} and Wide-field Infrared Survey Explorer
(WISE)\citep{wise}.  We find 17 objects whose spectral energy
distribution is consistent with a black-body radiation to a high
accuracy in the UV (GALEX; FUV, NUV), the optical (SDSS; $u,g,r,i,z$,)
and in the infrared (WISE; W1). These stars all are $r>17$.
The physical identity of these objects is yet to be confirmed, but we find that
they are consistent with DB white dwarfs with a low temperature.  
Whatever is the nature of the objects, we emphasize a practical value of these
black-body-like objects as photometric or spectrophotometric standards.

Such an object was found serendipitously while the quasar spectra in
Baryon Acoustic Oscillation Spectroscopic Survey of SDSS-III
\citep[BOSS,][]{dawson2013} were visually inspected.  This object
was primarily targeted as a quasar candidate \citep{ross2012},
but it does not resemble any type of quasars, and was eventually
rejected from the quasar catalogue \citep{paris2014}. It turned out
to be a star with a smooth-featureless spectrum.  In fact, this
object shows a proper motion, which, together with the close
proximity in its spectrum, leads us to conclude that this is a star
having a black-body spectrum.

We have then conducted a systematic search for such stars showing the black-body-like spectrum from the
archived 4.3 million SDSS spectral objects.  We select 22 stellar-like objects showing smooth black-body-like spectra. We
then employ GALEX satellite data for ultraviolet photometry.  We study if a black-body spectrum in the optical  wavelengths continues to UV. Particularly, it is important to find the peak and turnover of the black-body spectrum in the UV region:
we drop one star that does not pass this test.
We finally employ WISE satellite data for
infrared photometry.  It is amazing to see that the majority of the
stars we found for black-body-like star candidates show spectra that
extrapolate smoothly to the infrared as far as 3.4$\mu$m.
It is known that some stars,
specifically white dwarfs, are occasionally surrounded by a disc or
dust debris that exhibits some IR excess \citep{debes2011}.  We carried
out this test: 17 objects survived, and we present these
objects as black-body stars in this paper.  We dropped four objects
that show an IR excess from our sample.

We may use these black-body-like stars for the photometric standard for the
verification of the photometric system across various photometric
passbands, to the accuracy that they are indeed black-body stars. This enables
us to examine photometric systems used in the literature, in
particular the accuracy of the zero points, across various
photometric bands, which otherwise cannot easily be carried out. This
also allows us to examine possible relative consistency among a few
photometric systems that adopt the `AB system' outside the optical
passband used in different projects.  These black-body stars can also
be used as standard stars for photometric and spectrophotometric
observations.  Our objects all range in 17$-$19th mag in the $r$
band, which is appropriate for standard stars for 8$-$10 metre
telescopes.

%% file: 02_data.tex
\section{Data and Search for Black-Body Stars}
\subsection{SDSS}
 Our primary source is the spectroscopic data archive of SDSS in
DR7 \citep{DR7} and DR12 \citep{DR12}.  There are little
 overlaps
in targets between the two data sets, while the spectrograph
 was
modified between the two.  After DR7, the final data release of the SDSS I and II
projects, the spectrograph has been
modified for SDSS-III: the wavelength range has widened, the fibre
aperture size was narrowed, and washers are installed to correct for a
tilt against
 the focal plane in the spectrograph to enhance the
signal-to-noise ratio in blue wavelengths, while sacrificing the
quality of the flux calibration \citep{dawson2013}.

In the SDSS data reduction spectrophotometric flux is integrated
over
 the filter curve and calibrated against its broad-band,
point-spread-function (PSF) photometry of stars. 
We confirmed that spectrophotometric flux agrees with broad-band 
photometry within a 1\% systematic difference with an rms scatter, 
however, of 7\% in the $g-$, $r-$ and $i$-bands.
We remark here that the precision of broadband photometry is typically
1\% for the $g,r,i,z$-bands and 2\% for the $u$-band, as claimed in
the DR7 SDSS
 document.  Spectroscopic flux is then calibrated
against the model of
 \citet{gray2001} for
spectrophotometric standard stars per
 spectroscopic plate, typically
F subdwarfs, as described in the SDSS
 DR6 paper \citep{DR6}.

The observed spectra are automatically classified into object types by the
SDSS pipeline \citep{bolton2012}. The majority of the objects are
correctly classified, but classification into `quasars', often
suffers from contaminations of other types of objects, or sometimes redshift
 mismeasured, by a few percent probability.  Visual inspection is
conducted for all spectra classified as \lq\lq Quasar".

Quasars and white dwarfs often have similar colours, resembling the
black-body spectrum. Hence, many white dwarfs and the targets we are
looking for are included in the candidate quasar sample in the
target selection of SDSS. We go back to the master data archives.
The
 essential initial step of our work is to search for
featureless
 spectra with rather close to black-body in the data
archive of
 SDSS-I/II \citep[Data Release 8,][]{DR8} and SDSS-III
\citep[Data
 Release 12,][]{DR12}.  We use the data reduction version
of {\rm v5\_3\_12} for DR8, and {\rm v5\_7\_0} for DR12.  We
examine 1,843,200 spectra in DR8 and 2,463,000 spectra in DR12.  We
fit the black-body spectrum to all spectra with two free parameters,
the effective temperature and a flux normalization factor.  We did not
introduce reddening or a spectral tilt at this stage.  We
preselect spectra with $S/N> 20$ per pixel at the $r$-band wavelength
to avoid noisy spectra
 that would obscure absorption features.  We
collect objects whose reduced $\chi^{2}/{\rm dof}$ of the black-body
fit is less than 1.05.  We find 465 objects from the total of 798,593
star-like objects.  We find that the effective temperature is roughly
10,000K for most of the objects we are looking for.  Then, we visually
inspected all 465 objects and rejected ones that show
absorption features.  We remark that DC white dwarf is defined as a star
with featureless spectrum with the strength of the absorption lines
less than 5$\%$ of its continuum.
We rejected spectra with H or He absorption lines if they
are visible, if weak, within the allowed S/N of the SDSS data. 22 stars
survived this selection.

Once we identify candidates for featureless objects with spectra
close to the black-body in the spectral archives, we employ the five
broad-band photometry data from SDSS DR8 to enhance the
photometric accuracy in the optical band.  We remark that DR8
presents the last release of photometry of SDSS and includes all
SDSS objects\footnote{The data in DR8 occasionally differ from those
in DR7 due to the re-reduction the  data by the SDSS team.
  The difference between the two data releases is typically
of the order of 0.01 mag in either positive and negative way, but
it amounts for some cases to 0.05 mag. The difference does not seem
to show any systematic trends}. After confirming that the
photometric data are consistent with spectrophotometry, we refit
to obtain the temperature and the flux normalization from broad-band photometry,
whereas we should wait for the introduction of the GALEX UV data to 
give more definitive temperatures.

SDSS photometry is designed to be in the AB magnitude system. It
is constructed based on the absolute flux of $\alpha$ Lyr (Vega), and
used spectrophotometry of BD+17$^\circ$4708 as the intermediary.
Slight offsets, however, are claimed in order to make it closer to
the AB system in SDSS photometry.  For example it is suggested in
their data release paper (say DR8) that some offset be added to the
values given in the SDSS photometry data release for DR7 and DR8:
AB$-$SDSS= $-$0.042, $+$0.036, $+$0.015, $+$0.013 and $-$0.002
\footnote{These offsets are being used in the SDSS pipeline routine named ``SDSSFLUX2AB"
 in their IDL package.  It was derived by David Hogg (2003, private communication).}
\footnote{IDL package can be found in :
  \url{https://users.obs.carnegiescience.edu/yshen/IDL/photoop_doc.html}}
for $u,g,r,i,z$-bands from SDSS magnitude to (pseudo)AB magnitude.
It is one of our purposes to 
examine the validity of this possible offset, as seen in the next
section.

\subsection{GALEX}

We extend our study to include the UV using the all-sky survey of
GALEX \citep{galex}. We fit a black-body spectrum to these spectral
energy distributions (SED) in addition to those of 
SDSS, both taking broad-band photometry, and examine if the fit is 
consistent with black-body fitted to the optical band alone.

We collect the photometric data for FUV ($\lambda$1350-1780) and NUV
($\lambda$1770-2730). GALEX photometry has been calibrated to
STScI CALSPEC \citep{bohlin2001} in its AB magnitude system.  GALEX
photometry is important for our project to find the turnover of
black-body spectra when temperature is $<20,000$ K, which is indicated
in our fit to the optical spectrum. One candidate star,
J121886+414800, does not show a turn over in UV, so is dropped from
our black-body star sample. This makes our sample 21 stars.
We derive the effective temperature and
flux normalization
 factor from fit to the photometric data of SDSS
and GALEX, which
 are shown in Table 1 below.  We reserve for a
possibility in mind that the zero point of the
 AB magnitude of GALEX
may deviate from that of the SDSS AB photometric system.

We also check for the proper motion of black-body candidate stars
\citep{propermotion}.  We should reject candidates when they exhibit no
proper motions. They might be candidates for BL Lac objects. 
All our 22 candidate objects, however, show significant proper motions of the order of
20$-$70 mas/yr per coordinate, which is consistent with their
distances tens of parsec. This distance is consistent with distance
moduli inferred from brightness and general luminosity if these stars 
are white dwarf like objects.  In any case, we conclude that these stars are
in the local field.

\subsection{WISE}

We supplement our study with infrared photometry of WISE, which has
observed the sky at 3.4, 4.6, 12 and 22 $\mu$m (W1, W2, W3 and W4
filters, respectively) \citep{wise}.  We use the photometric
catalogue
 of \citet{lang2016}, which introduced ``forced'' photometry, 
where the positions of objects are fixed to those in SDSS
photometry, and unblurred images are coadded (unWISE) to deepen the
survey depth\footnote{We thank Brice M\'enard for his suggestion to use
 forced photometry to increase the depth to match our study.}.
Only the W1 (3.4$\mu$m) channel is appropriate to our work, for photometry matches
the depth we require only in this passband.  WISE photometry is
presented taking $\alpha$ Lyr (Vega) as the zero magnitude. 
This is transformed to the AB system using the conversion given in
\citet{wise}. We reserve here again for
 a possibility of mismatch in
the zero point of AB photometry between
 WISE and SDSS.  We remark
that the original WISE data release does not
 reach the depth we
 need.

It has been known that a
significant fraction (say, 20\%) of white
 dwarfs have dust debris
around the star that is manifested as an IR
 excess (Debes et al
2011).  The IR data are essential to rule out
 objects with the IR
excess. We find that 4 among 22 objects show a significant IR excess, and drop
them from our sample of black-body stars. This makes our
 final
sample to be 17. For this sample, the black-body fit from the UV
 to
the optical region extrapolates well to 3.4$\mu$m.

%% file: 03_result.tex
\section{Result}


Our initial 22 black body star candidates are presented in Table 1. It
includes 4 stars showing the IR excess and one star that does not
fit GALEX UV.  We carry out global black-body fits to the available broad-band photometric data
of SDSS and GALEX.  
The fit parameters are $T_{\rm eff}$ and the flux normalization factor $a$, as:
\begin{equation}
f_{\lambda}=a\frac{2hc^{2}}{\lambda^{5}}\frac{1}{\exp(hc/\lambda kT_{\rm eff})-1}
\end{equation}  
where $\lambda$ is the wavelength, $h$ the Planck constant and $k$ is the Boltzmann constant.  
The normalization $a$ is a dimensionless factor and given in this table as well as
$\chi^2$/dof.
We do not introduce extinction corrections in this fit.
 The normalization factor is alternatively represented as
the angular size $\theta$,
\begin{equation}
\theta={R \over d}=\sqrt{f_\nu(T_{\rm eff}) \over \pi B_\nu(T_{\rm eff})}=\sqrt{a \over \pi},
\end{equation}
where $B_\nu({\rm T_{eff}})$ is the Planck radiation brightness, $R$ is the
radius, and $d$ is the distance to the star.

In Figure 1 we present the spectra of the 22 stars given in Table 1
and the fits to the black-body spectrum. We include one star with its
UV deviated from GALEX.  The fit is carried out without using WISE
photometry as a constraint. This then shows that the IR excess is
apparent for four stars beyond 2 sigma.  When an IR excess is not
apparent, the fit from optical spectra extrapolates well to IR
3.4$\mu$m (within 1.2 sigma), which corroborates
 the conclusion that
the emission from the stars is close to black-body.  To scrutinize the
deviation from the black-body spectrum, we show in
 the lower panels
the residual of the observed flux against the black-body fit given
in the fraction. 
For WISE W1 we show this separately, as its error is too large
to display in the same scale as other bands.
We are left with 17 black-body stars.
Note that chi squares are not necessarily good for some stars, so that
one may
further reject some depending on their purpose.

We note that the temperature and the normalization are correlated in
the fit. We demonstrate an example for J124535.626$+$423824.58 in Figure 2,
where thick-hatched region is the 1 $\sigma$ allowed range in the fit to the SDSS
data only. This is squeezed to a thicker-hatched region upon the use
of the GALEX data, resulting in the reduced temperature error of $\pm 50$K. The
inclusion of WISE does not improve the result. This trend differs little
for the other 16 stars.

The black-body fit and the residuals are shown in
Table 2: the residuals are also plotted in Figure 3(a).
Brightness given in Table 2, assuming that the emission
is perfect black-body, may be used as the
photometric standard, till more accurate observation becomes available.
We added our calculation for the $J$, $H$, and $K$ bands
of Subaru MOIRCS \citep{suzuki2008} and W1 for WISE. 
NIR is not used as a constraint to the fit.
The zero-point is SDSS pseudo-AB system.
The mean residuals over our 17 stars are given in the upper row of Table 3.  For the optical
bands, they are all consistent with zero within errors of
0.01 mag, smaller than the typical error of $u$ or $z$ band
photometry.  This means that the SDSS photometry zero points are
consistent across the $u$ to the $z$ bands: no appreciable deviations are
seen in the zero-point of any of the five passbands.  We also see that
deviations are of random nature without showing systematic trends.
Looking at the deviations star-by-star, there seems no apparent
correlations between colour bands.  The same can be said for GALEX
NUV.  There seems no offset in the GALEX zero point against the SDSS
AB system beyond the photometric scatter.  The deviation 
is recognizable for GALEX FUV: observed brightness is fainter by
0.1-0.2 mag than
the black-body, while the large photometric scatter does not allow us
to draw a precise conclusion.
For W1 of WISE, the offset of mean is 0.452$\pm$0.663. The error is
large, while it is consistent with zero.  
We note that we use forced photometry of WISE 
and we do not use WISE W1 as a constraint to fit to the black-body.
We cannot derive an accurate conclusion as to the zero point of WISE photometry.

It has been alleged that SDSS photometry is deviated from the AB
system and some offsets should be added to the bare value of SDSS
photometric pipeline output\footnote{This has been claimed by David
  Hogg (private communication) and is used in some analyses, in
  particular of spectrophotometric data handling, of SDSS.}.  We
compare SDSS photometry to that with the suggested offsets (as given
in the bottom of Sect. 2.1) added to the data, and re-minimize the
black-body fit that is given in row 3 of Table 3. This is also shown
in panel (b) of Figure 3.  We now see the residuals increased
significantly, in particular in the $u$ band, where a rather large
offset ($-0.042$ mag) is added.  The residual to the black-body seen
in this figure looks somewhat larger and also wiggly across the five
colour bands: the observed is brighter in $u$ and fainter in $g$.

We also compare SDSS photometry to photometry
with the AB zero point set by CALSPEC \citep{holzman2008}, by
re-minimizing again the entire fit, as shown in row 5 in Table
3\footnote{The constant added is $-0.037$ ($u$), 0.024 ($g$), 0.005
  ($r$), 0.018 ($i$) and $-0.016$ ($z$)}, and also in panel (c) of Figure 3.
We observe a wiggle similar
to the plot with Hogg's offset above, indicating brighter $u$ and
fainter $g$.  Departures of residuals from zero are larger in the $u$
and $z$ bands, 0.038 mag, and $-$0.027 mag, respectively, which are
compared to $<0.01$ mag offsets with the original magnitude system of
SDSS.  We conclude that original SDSS photometry works better than
adding some offsets or using the CALSPEC standard as the AB zero
magnitude.

We see in these Figure and Table that GALEX NUV photometry smoothly
matches SDSS photometry, to the level of a few times 0.01 mag, albeit
the scatter of photometry in the NUV band is as large as 0.08 mag,
verifying the consistency between the SDSS photometric system and
GALEX NUV.  For FUV photometry a large scatter of the order of 0.1 mag
seen in our residual figure hinders us from deriving an accurate
conclusion as to the zero point offset, but the mismatch of the AB
magnitudes between the two systems is at most of the order of 0.2 mag.
Our finding here indicates that these stars can be used as
spectrophotometric standards from UV to IR with expected errors and
fluctuations in mind.

Our fit gives the temperature of our black-body stars 8500
to 12000K. 
So we may conclude that these stars are consistent with DB white
dwarfs with He lines undeveloped. The white dwarf spectrum is
generically not very far from the black-body spectrum up to absorption
features.  If we assume that these black-body stars are white dwarfs,
the observed brightness indicates the distance to be roughly of the
order of 50 pc. This is also consistent with the distance indicated by
large proper motions. A typical distance of 50 pc in the Galactic disc
indicates extinction $E(B-V)\approx 0.01$ or $A_r\approx 0.03$ mag.
Note that we cannot estimate the distance from colours.
We also note that dust reddening is almost parallel to the temperature in
colour space.  The reddening correction modifies the resulting
temperature, but, as we confirmed, the residuals of the black-body fits
are modified very little, only up to 0.01 mag even for a very large extinction,
$E(B-V)=0.10$.  Therefore, our conclusion as to black-body stars
is not modified.  The fitted temperature becomes lower as $\Delta
T\approx 122[E(B-V)/0.01]K$ upon the inclusion of reddening.

%% file: 04_conclusion.tex
\section{Conclusion} 

We have discovered 17 stars that show
spectra very close to the
 black-body for a wide range of spectrum
from GALEX FUV to the IR
 $\lambda=3.4\mu$m band out of 4,300,000
spectra archived in the SDSS
 data base. These spectra can give us a
unique possibility to examine the
 magnitude system, in particular
the zero point of the photometric
 system in AB, across the various colour
bands within SDSS, or even across
 the different system used by GALEX
against that of SDSS, or WISE
 versus SDSS.  Within the SDSS
photometric system, we would indicate that
 the zero point of the 5
colour bands is consistent within 0.01 mag.
We do not find indication for the offset
 with SDSS AB photometry up to
the absolute constant common in the 5 bands. We
do {\it not} need
 to introduce any additional constant to adjust to the
`AB system'.
 GALEX NUV photometry zero point is consistent with SDSS
photometric
 zero point within a few hundredths of mag, although
large scatter of NUV photometry ($\approx$0.07 mag) hinders from a more
accurate conclusion.  During this study we also noted a 0.02$-$0.04
mag error in the optical zero point of the CALSPEC AB standard.
 
The black-body stars we found are as faint as 17$-$19 mag in $r$, and
can be used as photometric, but also as spectrophotometric standard
stars for the work with large aperture telescopes.  We consider that
these stars in our sample are consistent with DB white dwarfs with
temperature around 10000 K.

\noindent{\bf Acknowledgment}

We thank Brice M\'enard for useful discussion and Ting-Wen Lan for
assisting us to collect the WISE data.
NS is partially supported by JST CREST JPMHCR1414 and JSPS Programs 
for Advancing Strategic International Networks to Accelerate the Circulation of
Talented Researchers.
MF thanks Hans B\"ohringer and Yasuo Tanaka for
the hospitality at the Max-Planck-Institut f\"ur Extraterrestrische
Physik and also Eiichiro Komatsu at Max-Planck-Institut f\"ur
Astrophysik, in Garching. He also wishes his thanks to Alexander von
Humboldt Stiftung for the support during his stay in Garching, and
Monell Foundation in Princeton.  He received in Tokyo a Grant-in-Aid (No. 154300000110) from the Ministry
of Education.  Kavli IPMU is supported by World Premier International
Research Center Initiative of the Ministry of Education, Japan.

%% file: table_a.tex
\begin{landscape}
\begin{deluxetable}{lrrrrrrrrrrrrrrrr}
\tablewidth{9.3in}
\tabletypesize{\tiny}
\tablecaption{22 Black-body Candidates: Coordinates and Photometric Data}
\tablehead{
\colhead{SDSS name}&
\colhead{RA (J2000)}&
\colhead{DEC (J2000)}&
\colhead{GALEX FUV}&
\colhead{GALEX NUV}&
\colhead{SDSS $u$}&
\colhead{SDSS $g$}&
\colhead{SDSS $r$}&
\colhead{SDSS $i$}&
\colhead{SDSS $z$}&
\colhead{WISE W1}&
\colhead{T (K)}&
\colhead{$a$($\times10^{-23}$)}&
\colhead{$\theta$("$\times10^{-9}$)}&
\colhead{$\chi^{2}$/dof}\\
}
\label{tbl:bbfit}
\startdata
J002739.497-001741.93&00:27:39.497   &-00:17:41.93   &  21.80$\pm$   0.06&  19.89$\pm$   0.02&  19.05$\pm$   0.02&  18.90$\pm$   0.03&  18.98$\pm$   0.02&  19.12$\pm$   0.03&  19.37$\pm$   0.05&  21.21$\pm$   0.48&  10662$\pm$     60& 0.435$\pm$ 0.010& 242.7$\pm$   2.8& 0.84\\
J004705.818-004820.09\tablenotemark{*}&00:47:05.818   &-00:48:20.09   &  20.02$\pm$   0.04&  19.29$\pm$   0.02&  19.00$\pm$   0.03&  19.01$\pm$   0.02&  19.24$\pm$   0.02&  19.50$\pm$   0.02&  19.60$\pm$   0.05&  21.07$\pm$   0.41&  14648$\pm$     94& 0.164$\pm$ 0.003& 149.0$\pm$   1.6& 8.67\\
J004830.324+001752.80&00:48:30.324   &+00:17:52.80   &  21.18$\pm$   0.05&  19.14$\pm$   0.01&  18.27$\pm$   0.03&  18.14$\pm$   0.01&  18.23$\pm$   0.01&  18.38$\pm$   0.01&  18.63$\pm$   0.03&  20.46$\pm$   0.22&  10639$\pm$     40& 0.876$\pm$ 0.013& 344.5$\pm$   2.5& 5.10\\
J014618.898-005150.51&01:46:18.898   &-00:51:50.51   &  20.61$\pm$   0.05&  18.80$\pm$   0.01&  18.21$\pm$   0.02&  18.17$\pm$   0.02&  18.24$\pm$   0.01&  18.42$\pm$   0.01&  18.61$\pm$   0.03&  21.19$\pm$   0.41&  11770$\pm$     48& 0.672$\pm$ 0.010& 301.8$\pm$   2.3& 8.52\\
J022936.715-004113.63&02:29:36.715   &-00:41:13.63   &  23.04$\pm$   0.14&  20.73$\pm$   0.01&  19.38$\pm$   0.03&  19.07$\pm$   0.03&  19.09$\pm$   0.01&  19.17$\pm$   0.02&  19.31$\pm$   0.04&  21.00$\pm$   0.30&   8901$\pm$     31& 0.640$\pm$ 0.011& 294.5$\pm$   2.5& 1.93\\
J083226.568+370955.48&08:32:26.568   &+37:09:55.48   &&&  19.39$\pm$   0.03&  18.98$\pm$   0.01&  18.84$\pm$   0.03&  18.83$\pm$   0.02&  18.93$\pm$   0.04&  20.63$\pm$   0.24&   7952$\pm$     95& 1.137$\pm$ 0.050& 392.4$\pm$   8.5& 0.63\\
J083736.557+542758.64&08:37:36.557   &+54:27:58.64   &&  21.18$\pm$   0.07&  19.24$\pm$   0.03&  18.73$\pm$   0.03&  18.56$\pm$   0.02&  18.54$\pm$   0.02&  18.52$\pm$   0.04&  20.34$\pm$   0.16&   7449$\pm$     61& 1.815$\pm$ 0.055& 495.8$\pm$   7.5& 3.38\\
J100449.541+121559.65&10:04:49.541   &+12:15:59.65   &&&  19.41$\pm$   0.02&  19.18$\pm$   0.03&  19.23$\pm$   0.02&  19.32$\pm$   0.02&  19.43$\pm$   0.05&  22.25$\pm$   1.07&   9773$\pm$    136& 0.440$\pm$ 0.019& 244.1$\pm$   5.2& 0.57\\
J103123.906+093657.89\tablenotemark{*}&10:31:23.906   &+09:36:57.89   &&&  18.73$\pm$   0.03&  18.64$\pm$   0.02&  18.80$\pm$   0.01&  18.96$\pm$   0.02&  19.15$\pm$   0.05&  20.63$\pm$   0.25&  11667$\pm$    224& 0.419$\pm$ 0.020& 238.2$\pm$   5.7& 0.40\\
J104523.866+015721.96&10:45:23.866   &+01:57:21.96   &&  20.51$\pm$   0.09&  19.32$\pm$   0.03&  19.09$\pm$   0.01&  19.01$\pm$   0.01&  19.07$\pm$   0.02&  19.28$\pm$   0.07&  22.13$\pm$   1.04&   8956$\pm$     87& 0.666$\pm$ 0.021& 300.3$\pm$   4.7& 2.56\\
J111720.801+405954.67&11:17:20.801   &+40:59:54.67   &  20.93$\pm$   0.34&  18.98$\pm$   0.09&  18.26$\pm$   0.02&  18.08$\pm$   0.01&  18.19$\pm$   0.01&  18.34$\pm$   0.01&  18.63$\pm$   0.04&  23.00$\pm$   1.91&  10950$\pm$    121& 0.843$\pm$ 0.026& 337.9$\pm$   5.1& 2.56\\
J114722.608+171325.21&11:47:22.608   &+17:13:25.21   &&  19.91$\pm$   0.06&  18.91$\pm$   0.03&  18.65$\pm$   0.02&  18.68$\pm$   0.02&  18.85$\pm$   0.02&  19.00$\pm$   0.04&  22.24$\pm$   1.06&   9962$\pm$    109& 0.669$\pm$ 0.022& 301.0$\pm$   4.9& 1.92\\
J121856.693+414800.29 \tablenotemark{a}&12:18:56.693   &+41:48:00.29   &  17.59$\pm$   0.06&  17.84$\pm$   0.04&  17.91$\pm$   0.02&  18.18$\pm$   0.02&  18.62$\pm$   0.02&  18.90$\pm$   0.01&  19.22$\pm$   0.05&  22.58$\pm$   1.26&  23357$\pm$    383& 0.133$\pm$ 0.004& 134.1$\pm$   2.0& 7.03\\
J124535.626+423824.58&12:45:35.626   &+42:38:24.58   &&  18.30$\pm$   0.03&  17.32$\pm$   0.02&  17.14$\pm$   0.02&  17.18$\pm$   0.02&  17.29$\pm$   0.01&  17.46$\pm$   0.02&  19.60$\pm$   0.08&  10086$\pm$     67& 2.650$\pm$ 0.057& 599.1$\pm$   6.4& 0.17\\
J125507.082+192459.00&12:55:07.082   &+19:24:59.00   &&  19.93$\pm$   0.08&  18.83$\pm$   0.04&  18.53$\pm$   0.02&  18.45$\pm$   0.01&  18.50$\pm$   0.01&  18.63$\pm$   0.03&  21.75$\pm$   0.78&   8882$\pm$     98& 1.157$\pm$ 0.039& 395.8$\pm$   6.6& 1.46\\
J134305.302+270623.98&13:43:05.302   &+27:06:23.98   &&  20.03$\pm$   0.14&  19.05$\pm$   0.02&  18.93$\pm$   0.02&  19.00$\pm$   0.02&  19.14$\pm$   0.02&  19.34$\pm$   0.07&  22.09$\pm$   0.85&  10678$\pm$    151& 0.427$\pm$ 0.017& 240.4$\pm$   4.8& 0.20\\
J135816.735+144202.27 \tablenotemark{*}&13:58:16.735   &+14:42:02.27   &  21.93$\pm$   0.12&  19.46$\pm$   0.02&  18.34$\pm$   0.03&  18.06$\pm$   0.03&  18.02$\pm$   0.02&  18.07$\pm$   0.01&  18.12$\pm$   0.03&  19.77$\pm$   0.09&   9191$\pm$     44& 1.598$\pm$ 0.030& 465.2$\pm$   4.3& 3.39\\
J141724.329+494127.85&14:17:24.329   &+49:41:27.85   &  20.73$\pm$   0.29&  18.27$\pm$   0.06&  17.36$\pm$   0.03&  17.25$\pm$   0.03&  17.31$\pm$   0.02&  17.43$\pm$   0.02&  17.61$\pm$   0.03&  20.01$\pm$   0.08&  10503$\pm$    112& 2.127$\pm$ 0.066& 536.7$\pm$   8.3& 1.01\\
J151859.717+002839.58&15:18:59.717   &+00:28:39.58   &&&  19.71$\pm$   0.03&  19.44$\pm$   0.01&  19.37$\pm$   0.02&  19.51$\pm$   0.03&  19.57$\pm$   0.07&  21.57$\pm$   0.47&   9072$\pm$    131& 0.458$\pm$ 0.022& 249.1$\pm$   5.9& 0.71\\
J161605.194+142116.70 \tablenotemark{*}&16:16:05.194   &+14:21:16.70   &  22.32$\pm$   0.47&  19.57$\pm$   0.07&  18.70$\pm$   0.02&  18.61$\pm$   0.01&  18.70$\pm$   0.01&  18.85$\pm$   0.02&  19.02$\pm$   0.04&  19.52$\pm$   0.07&  10918$\pm$    119& 0.533$\pm$ 0.016& 268.6$\pm$   4.1& 1.08\\
J161704.078+181311.96&16:17:04.078   &+18:13:11.96   &&  20.60$\pm$   0.12&  19.17$\pm$   0.03&  18.78$\pm$   0.02&  18.73$\pm$   0.02&  18.80$\pm$   0.02&  18.81$\pm$   0.04&  21.56$\pm$   0.44&   8568$\pm$     88& 0.996$\pm$ 0.034& 367.2$\pm$   6.2& 2.00\\
J230240.032-003021.60&23:02:40.032   &-00:30:21.60   &  20.79$\pm$   0.06&  18.88$\pm$   0.01&  17.97$\pm$   0.02&  17.80$\pm$   0.02&  17.90$\pm$   0.02&  18.02$\pm$   0.02&  18.25$\pm$   0.03&  21.19$\pm$   0.42&  10478$\pm$     42& 1.241$\pm$ 0.021& 409.9$\pm$   3.4& 1.14
\enddata
\tablenotetext{*}{Stars that show IR excess. To be dropped from the list of black-body stars.}
\tablenotetext{a}{GALEX UV data do not show turn-off. To be removed from the list of black-body stars.}
\end{deluxetable}
\end{landscape}

%% file: table_b.tex
\begin{landscape}
\begin{deluxetable}{lrrrrrrrrrrrr}
\tablewidth{9.1in}
\tabletypesize{\tiny}
\tablecaption{The best fit broad-band flux of the 17 black-body stars (upper rows) and the residuals from the fit (lower rows):$\Delta$m = Data$-$Black-Body in magnitude.  Calculations are added for Subaru MOIRCS J, H, K and for W1 of WISE.  NIR is not used as a constraint to the fit}
\label{tbl:deltam}
\tablehead{
\colhead{Star name}&
\colhead{GALEX FUV}&
\colhead{GALEX NUV}&
\colhead{SDSS-u}&
\colhead{SDSS-g}&
\colhead{SDSS-r}&
\colhead{SDSS-i}&
\colhead{SDSS-z}&
\colhead{$\chi^{2}$/dof}&
\colhead{MOIRCS-J}&
\colhead{MOIRCS-H}&
\colhead{MOIRCS-K}&
\colhead{WISE W1}
}
\startdata
J002739.497-001741.93&    21.705$\pm$     0.027&    19.905$\pm$     0.010&    19.041$\pm$     0.002&    18.911$\pm$     0.007&    18.980$\pm$     0.010&    19.124$\pm$     0.012&    19.306$\pm$     0.014&&    19.753$\pm$     0.015&    20.170$\pm$     0.016&    20.647$\pm$     0.017&    21.470$\pm$     0.018\\
&   0.090$\pm$   0.061&  -0.017$\pm$   0.019&   0.010$\pm$   0.022&  -0.013$\pm$   0.025&   0.000$\pm$   0.019&  -0.005$\pm$   0.034&   0.060$\pm$   0.034& 0.84& & &&   -0.259$\pm$   0.482\\
\hline
J004830.324+001752.80&    20.965$\pm$     0.019&    19.158$\pm$     0.008&    18.289$\pm$     0.000&    18.158$\pm$     0.003&    18.225$\pm$     0.006&    18.368$\pm$     0.007&    18.550$\pm$     0.008&&    18.996$\pm$     0.009&    19.413$\pm$     0.010&    19.889$\pm$     0.010&    20.712$\pm$     0.011\\
&   0.215$\pm$   0.049&  -0.021$\pm$   0.013&  -0.021$\pm$   0.027&  -0.017$\pm$   0.014&   0.007$\pm$   0.014&   0.007$\pm$   0.014&   0.082$\pm$   0.014& 5.10& & &&   -0.252$\pm$   0.217\\
\hline
J014618.898-005150.51&    20.351$\pm$     0.018&    18.839$\pm$     0.007&    18.169$\pm$     0.001&    18.121$\pm$     0.004&    18.250$\pm$     0.006&    18.425$\pm$     0.007&    18.630$\pm$     0.008&&    19.108$\pm$     0.010&    19.541$\pm$     0.010&    20.031$\pm$     0.011&    20.867$\pm$     0.011\\
&   0.263$\pm$   0.050&  -0.040$\pm$   0.012&   0.045$\pm$   0.019&   0.046$\pm$   0.019&  -0.006$\pm$   0.013&  -0.004$\pm$   0.015&  -0.025$\pm$   0.015& 8.52& & &&    0.327$\pm$   0.411\\
\hline
J022936.715-004113.63&    23.126$\pm$     0.020&    20.716$\pm$     0.007&    19.450$\pm$     0.001&    19.140$\pm$     0.005&    19.079$\pm$     0.008&    19.155$\pm$     0.010&    19.289$\pm$     0.011&&    19.668$\pm$     0.012&    20.049$\pm$     0.013&    20.498$\pm$     0.013&    21.293$\pm$     0.014\\
&  -0.086$\pm$   0.144&   0.009$\pm$   0.012&  -0.066$\pm$   0.029&  -0.065$\pm$   0.031&   0.008$\pm$   0.015&   0.012$\pm$   0.017&   0.025$\pm$   0.017& 1.93& & &&   -0.292$\pm$   0.302\\
\hline
J083226.568+370955.48&    23.824$\pm$     0.097&    20.972$\pm$     0.048&    19.420$\pm$     0.017&    18.977$\pm$     0.002&    18.819$\pm$     0.009&    18.844$\pm$     0.015&    18.942$\pm$     0.019&&    19.270$\pm$     0.024&    19.624$\pm$     0.027&    20.053$\pm$     0.030&    20.827$\pm$     0.032\\
&   &   &  -0.025$\pm$   0.031&   0.008$\pm$   0.015&   0.026$\pm$   0.028&  -0.017$\pm$   0.022&  -0.013$\pm$   0.022& 0.63& & &&   -0.201$\pm$   0.240\\
\hline
J083736.557+542758.64&    24.149$\pm$     0.073&    21.016$\pm$     0.037&    19.286$\pm$     0.014&    18.757$\pm$     0.003&    18.538$\pm$     0.005&    18.530$\pm$     0.009&    18.604$\pm$     0.012&&    18.899$\pm$     0.016&    19.235$\pm$     0.018&    19.651$\pm$     0.020&    20.412$\pm$     0.022\\
&   &   0.166$\pm$   0.075&  -0.045$\pm$   0.026&  -0.023$\pm$   0.027&   0.026$\pm$   0.017&   0.013$\pm$   0.017&  -0.081$\pm$   0.017& 3.38& & &&   -0.067$\pm$   0.162\\
\hline
J100449.541+121559.65&    22.541$\pm$     0.092&    20.461$\pm$     0.046&    19.411$\pm$     0.015&    19.199$\pm$     0.002&    19.209$\pm$     0.008&    19.322$\pm$     0.013&    19.482$\pm$     0.017&&    19.898$\pm$     0.022&    20.299$\pm$     0.025&    20.763$\pm$     0.027&    21.573$\pm$     0.029\\
&   &   &   0.004$\pm$   0.024&  -0.019$\pm$   0.025&   0.021$\pm$   0.023&   0.000$\pm$   0.021&  -0.048$\pm$   0.021& 0.57& & &&    0.679$\pm$   1.071\\
\hline
J104523.866+015721.96&    23.016$\pm$     0.071&    20.629$\pm$     0.036&    19.377$\pm$     0.013&    19.074$\pm$     0.002&    19.018$\pm$     0.005&    19.096$\pm$     0.009&    19.232$\pm$     0.012&&    19.614$\pm$     0.016&    19.996$\pm$     0.018&    20.446$\pm$     0.020&    21.242$\pm$     0.021\\
&   &  -0.118$\pm$   0.085&  -0.053$\pm$   0.028&   0.021$\pm$   0.010&  -0.005$\pm$   0.011&  -0.024$\pm$   0.016&   0.050$\pm$   0.016& 2.56& & &&    0.886$\pm$   1.038\\
\hline
J111720.801+405954.67&    20.741$\pm$     0.066&    19.021$\pm$     0.034&    18.211$\pm$     0.012&    18.105$\pm$     0.002&    18.190$\pm$     0.005&    18.343$\pm$     0.008&    18.531$\pm$     0.011&&    18.987$\pm$     0.014&    19.409$\pm$     0.016&    19.889$\pm$     0.018&    20.716$\pm$     0.019\\
&   0.186$\pm$   0.336&  -0.039$\pm$   0.086&   0.046$\pm$   0.022&  -0.027$\pm$   0.013&   0.002$\pm$   0.014&  -0.001$\pm$   0.014&   0.101$\pm$   0.014& 2.56& & &&    2.282$\pm$   1.910\\
\hline
J114722.608+171325.21&    21.893$\pm$     0.071&    19.877$\pm$     0.036&    18.869$\pm$     0.012&    18.676$\pm$     0.002&    18.699$\pm$     0.006&    18.819$\pm$     0.010&    18.985$\pm$     0.013&&    19.408$\pm$     0.017&    19.812$\pm$     0.019&    20.279$\pm$     0.020&    21.092$\pm$     0.022\\
&   &   0.036$\pm$   0.058&   0.043$\pm$   0.031&  -0.027$\pm$   0.016&  -0.016$\pm$   0.017&   0.035$\pm$   0.019&   0.017$\pm$   0.019& 1.92& & &&    1.152$\pm$   1.060\\
\hline
J124535.626+423824.58&    20.276$\pm$     0.042&    18.300$\pm$     0.020&    17.319$\pm$     0.006&    17.138$\pm$     0.000&    17.170$\pm$     0.005&    17.295$\pm$     0.007&    17.463$\pm$     0.009&&    17.891$\pm$     0.011&    18.297$\pm$     0.013&    18.767$\pm$     0.014&    19.581$\pm$     0.015\\
&   &  -0.004$\pm$   0.032&  -0.000$\pm$   0.017&   0.001$\pm$   0.017&   0.015$\pm$   0.019&  -0.007$\pm$   0.015&  -0.004$\pm$   0.015& 0.17& & &&    0.022$\pm$   0.081\\
\hline
J125507.082+192459.00&    22.508$\pm$     0.084&    20.090$\pm$     0.044&    18.818$\pm$     0.017&    18.506$\pm$     0.005&    18.443$\pm$     0.003&    18.518$\pm$     0.008&    18.651$\pm$     0.011&&    19.030$\pm$     0.016&    19.410$\pm$     0.018&    19.859$\pm$     0.020&    20.653$\pm$     0.022\\
&   &  -0.158$\pm$   0.081&   0.010$\pm$   0.038&   0.021$\pm$   0.019&   0.012$\pm$   0.015&  -0.015$\pm$   0.015&  -0.025$\pm$   0.015& 1.46& & &&    1.100$\pm$   0.781\\
\hline
J134305.302+270623.98&    21.712$\pm$     0.086&    19.916$\pm$     0.043&    19.055$\pm$     0.014&    18.927$\pm$     0.002&    18.996$\pm$     0.007&    19.141$\pm$     0.012&    19.323$\pm$     0.015&&    19.771$\pm$     0.019&    20.188$\pm$     0.022&    20.665$\pm$     0.024&    21.488$\pm$     0.026\\
&   &   0.117$\pm$   0.139&  -0.008$\pm$   0.024&  -0.002$\pm$   0.021&   0.005$\pm$   0.016&  -0.004$\pm$   0.020&   0.012$\pm$   0.020& 0.20& & &&    0.598$\pm$   0.846\\
\hline
J141724.329+494127.85&    20.124$\pm$     0.066&    18.277$\pm$     0.033&    17.382$\pm$     0.011&    17.238$\pm$     0.001&    17.297$\pm$     0.006&    17.436$\pm$     0.009&    17.615$\pm$     0.012&&    18.057$\pm$     0.015&    18.471$\pm$     0.017&    18.946$\pm$     0.019&    19.766$\pm$     0.020\\
&   0.608$\pm$   0.288&  -0.008$\pm$   0.056&  -0.026$\pm$   0.026&   0.012$\pm$   0.026&   0.010$\pm$   0.017&  -0.005$\pm$   0.022&  -0.006$\pm$   0.022& 1.01& & &&    0.242$\pm$   0.081\\
\hline
J151859.717+002839.58&    23.280$\pm$     0.102&    20.940$\pm$     0.050&    19.719$\pm$     0.017&    19.431$\pm$     0.001&    19.384$\pm$     0.010&    19.468$\pm$     0.016&    19.608$\pm$     0.020&&    19.995$\pm$     0.026&    20.380$\pm$     0.029&    20.832$\pm$     0.031&    21.630$\pm$     0.034\\
&   &   &  -0.007$\pm$   0.028&   0.007$\pm$   0.015&  -0.013$\pm$   0.016&   0.040$\pm$   0.033&  -0.042$\pm$   0.033& 0.71& & &&   -0.057$\pm$   0.473\\
\hline
J161704.078+181311.96&    23.078$\pm$     0.080&    20.524$\pm$     0.041&    19.164$\pm$     0.015&    18.811$\pm$     0.004&    18.719$\pm$     0.005&    18.778$\pm$     0.009&    18.900$\pm$     0.013&&    19.263$\pm$     0.017&    19.635$\pm$     0.020&    20.078$\pm$     0.021&    20.866$\pm$     0.023\\
&   &   0.074$\pm$   0.123&   0.006$\pm$   0.025&  -0.030$\pm$   0.021&   0.012$\pm$   0.016&   0.020$\pm$   0.017&  -0.092$\pm$   0.017& 2.00& & &&    0.691$\pm$   0.436\\
\hline
J230240.032-003021.60&    20.731$\pm$     0.020&    18.877$\pm$     0.007&    17.977$\pm$     0.001&    17.831$\pm$     0.005&    17.889$\pm$     0.007&    18.027$\pm$     0.009&    18.205$\pm$     0.010&&    18.646$\pm$     0.011&    19.060$\pm$     0.012&    19.534$\pm$     0.012&    20.354$\pm$     0.013\\
&   0.057$\pm$   0.056&   0.000$\pm$   0.013&  -0.009$\pm$   0.021&  -0.027$\pm$   0.017&   0.009$\pm$   0.017&  -0.002$\pm$   0.020&   0.041$\pm$   0.020& 1.14& & &&    0.838$\pm$   0.422
\enddata
\end{deluxetable}
\end{landscape}

%% file: table_c.tex
\begin{landscape}
\begin{deluxetable}{lrrrrrrr}
\tablewidth{7.5in}
\tabletypesize{\scriptsize}

\tablecaption{Mean of residuals from the black-body fit ($\Delta$m=data $-$ black-body fit): The 17 black-body stars are used in this calculation. Row 1 is with the raw value of brightness given in SDSS data release DR8; Row 2 gives the offsets suggested by Hogg to make the SDSS magnitude closer to the AB system, and row 3 is the mean residuals for this case.  Row 4  gives the offsets suggested by Holzman (2009) to make the SDSS magnitude closer to the AB system with CALSPEC, and Row 5 is the mean residuals.}
\tablehead{
\colhead{Data Name}&
\colhead{GALEX FUV}&
\colhead{GALEX NUV}&
\colhead{SDSS-u}&
\colhead{SDSS-g}&
\colhead{SDSS-r}&
\colhead{SDSS-i}&
\colhead{SDSS-z}\\
}
\label{tbl:deltam}
\startdata
Mean Residuals without Offset & 0.191$\pm$0.202&-0.000$\pm$0.081&-0.006$\pm$0.031&-0.008$\pm$0.025&0.007$\pm$0.012&0.002$\pm$0.017&0.003$\pm$0.052\\
\tableline
Hogg's Offset to SDSS&                &               & -0.042 & 0.036 & 0.015 & 0.013 & -0.002\\
Mean Residuals with Hogg Offset&  0.218$\pm$0.215& 0.016$\pm$0.097&-0.047$\pm$0.034&0.020$\pm$0.025&0.007$\pm$0.014&-0.002$\pm$0.016&-0.019$\pm$0.054\\
\tableline
Holtzman et al Offset (CALSEPC) &                &               & -0.037 & 0.024 & 0.005 & 0.018 & -0.016\\
Mean Residuals with Holtzman Offset &0.215$\pm$0.213&0.018$\pm$0.093&-0.038$\pm$0.033&0.013$\pm$0.025&0.003$\pm$0.013&0.009$\pm$0.016&-0.027$\pm$0.054
\enddata
\end{deluxetable}
\end{landscape}

%% file: figures.tex
\begin{figure}
\begin{center}
\hspace{-5mm}
\includegraphics[width=.49\linewidth,angle=0]{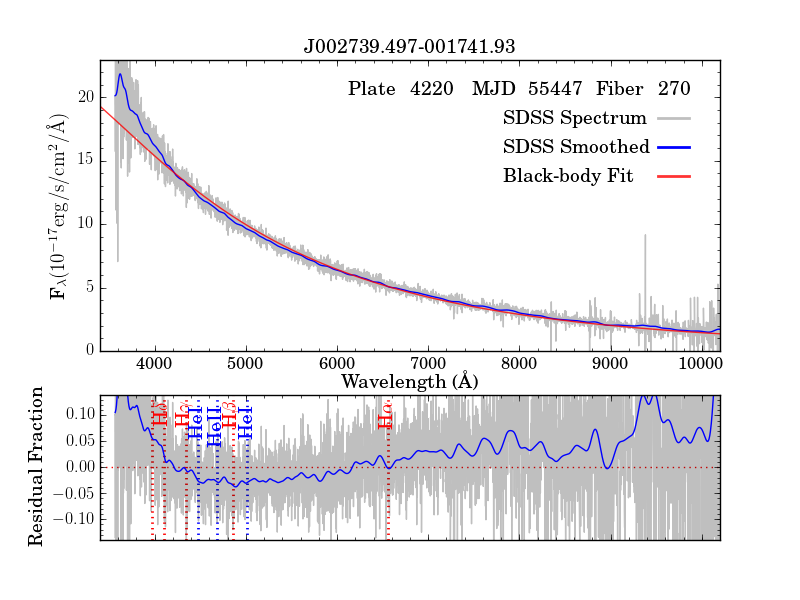}
\includegraphics[width=.49\linewidth,angle=0]{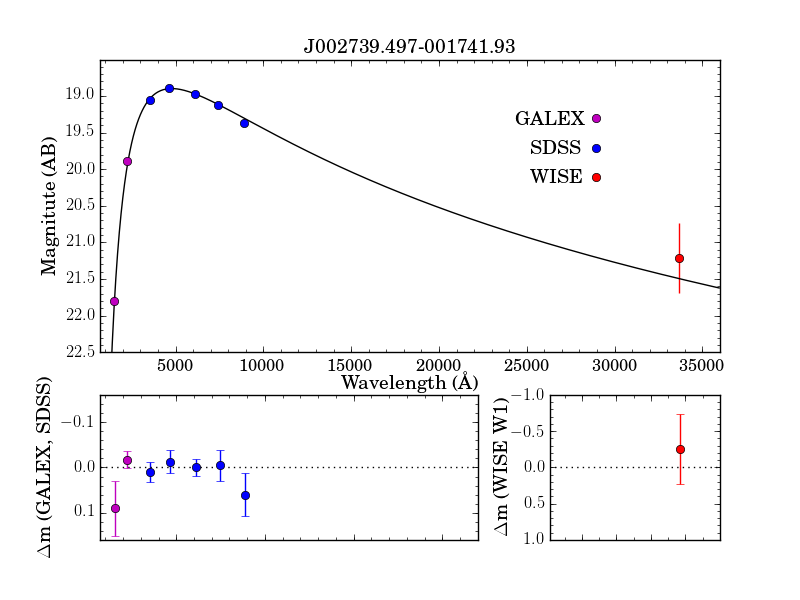}
\includegraphics[width=.49\linewidth,angle=0]{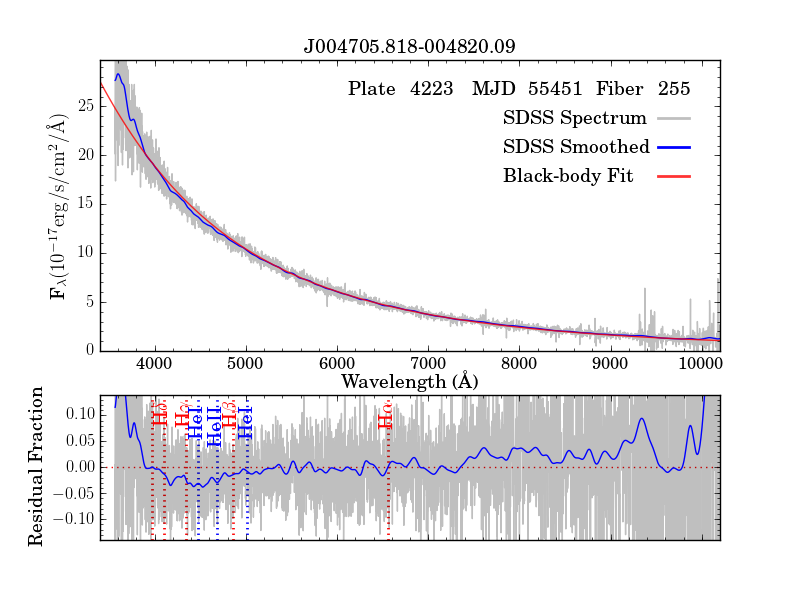}
\includegraphics[width=.49\linewidth,angle=0]{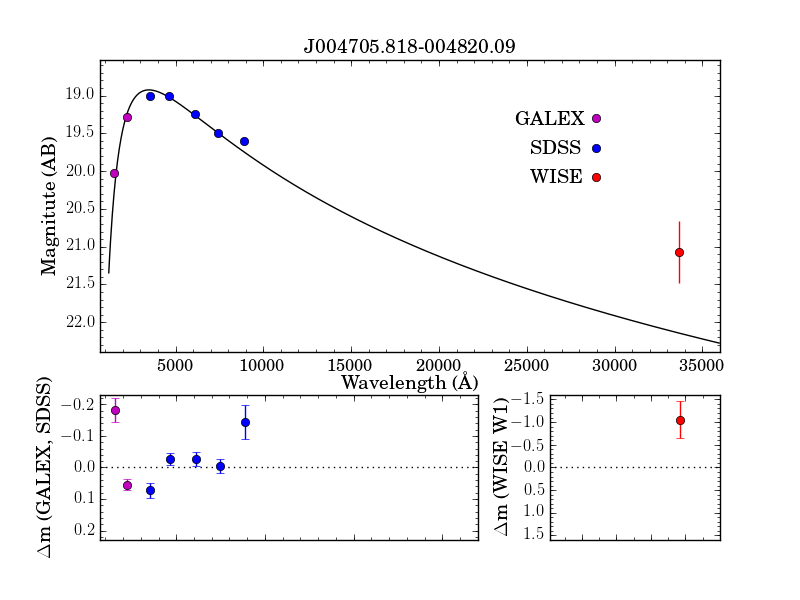}
\includegraphics[width=.49\linewidth,angle=0]{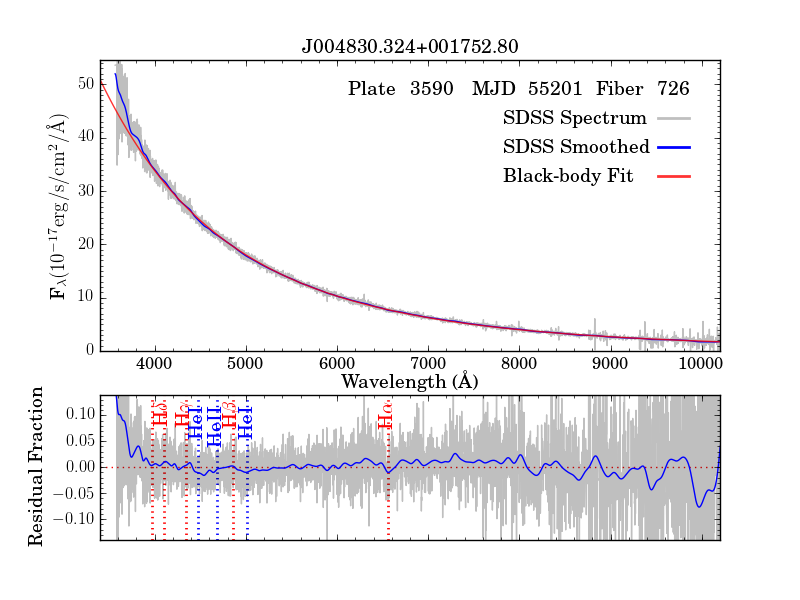}
\includegraphics[width=.49\linewidth,angle=0]{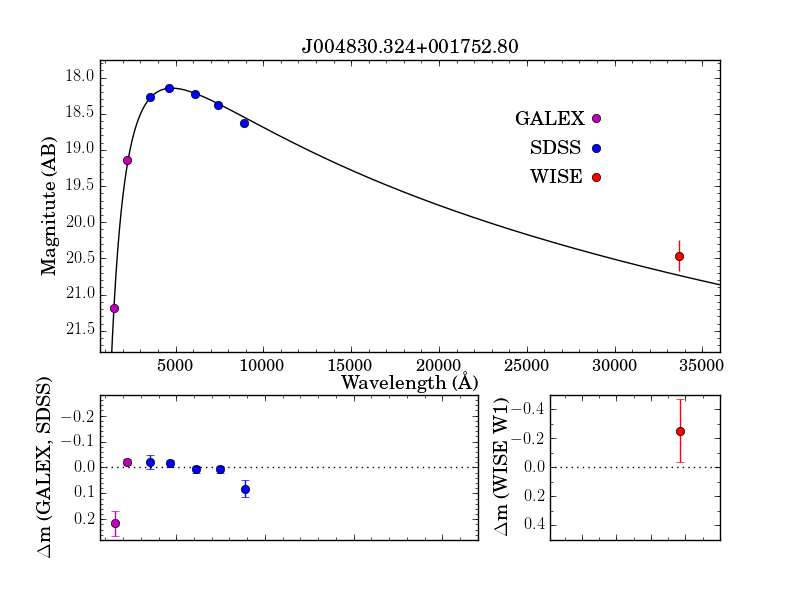}
\caption{
Spectra of 22 black-body star candidates.  In the left
panels, observed spectra are indicated with grey, and blue curves show
smoothed spectra. Black-body fits are indicated with red.  The right
panels show their photometric data from which our fit parameters are
derived. In the bottom panel, the residuals (fractional values) from
the fits are shown, with the positions of hydrogen and helium absorption
lines indicated.
 }
\end{center}
\end{figure}

\clearpage
\begin{figure}
\begin{center}
\includegraphics[width=.49\linewidth,angle=0]{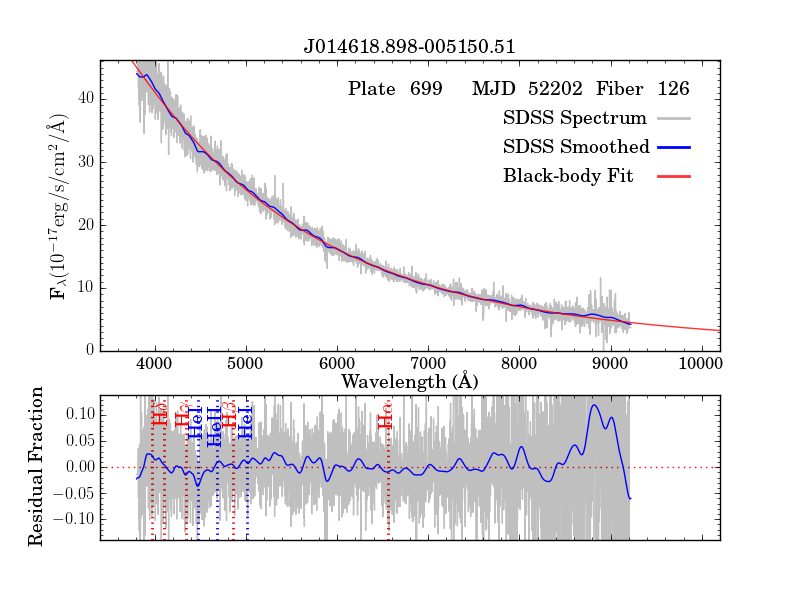}
\includegraphics[width=.49\linewidth,angle=0]{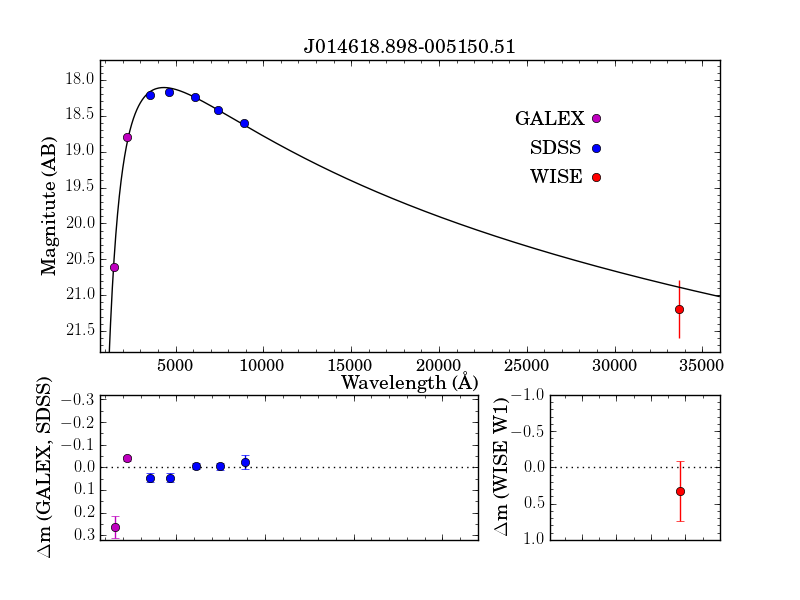}
\includegraphics[width=.49\linewidth,angle=0]{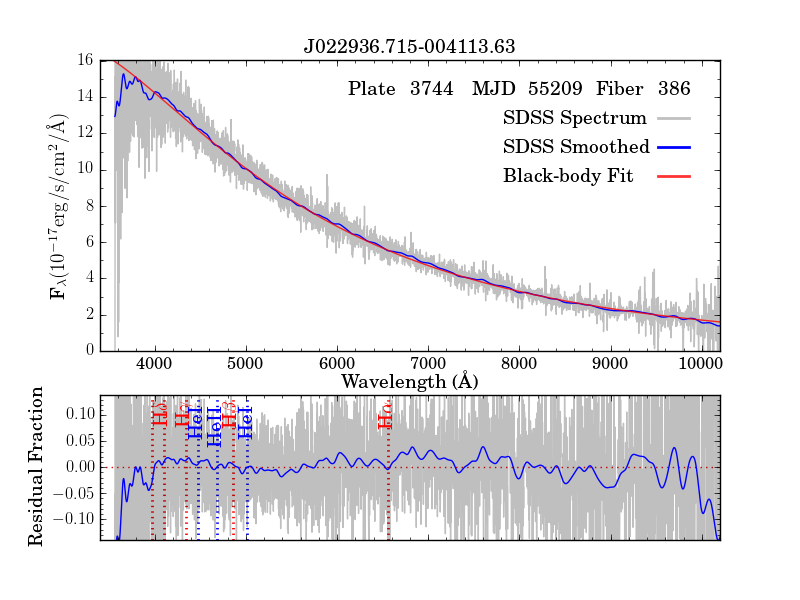}
\includegraphics[width=.49\linewidth,angle=0]{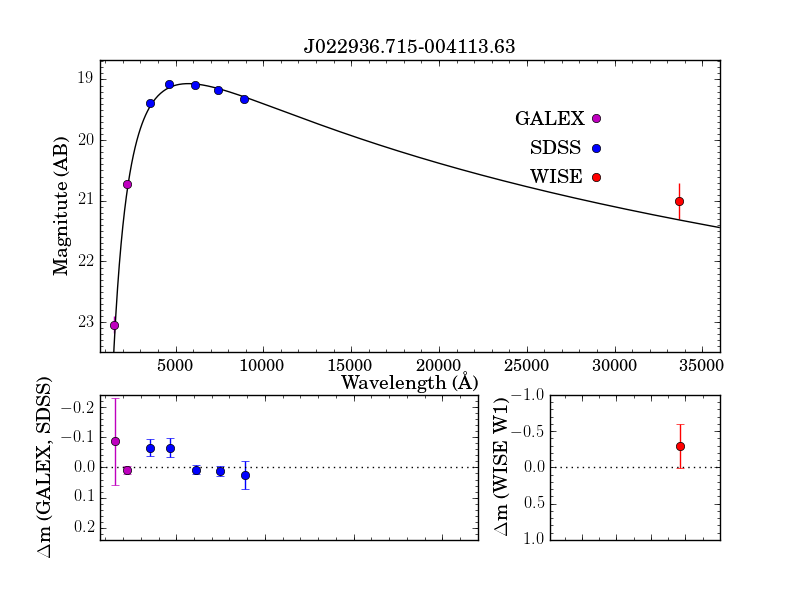}
\includegraphics[width=.49\linewidth,angle=0]{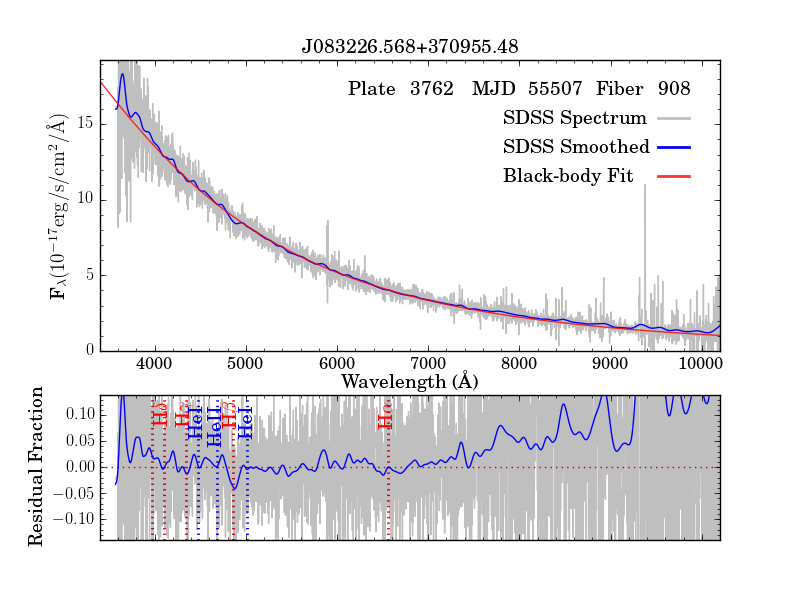}
\includegraphics[width=.49\linewidth,angle=0]{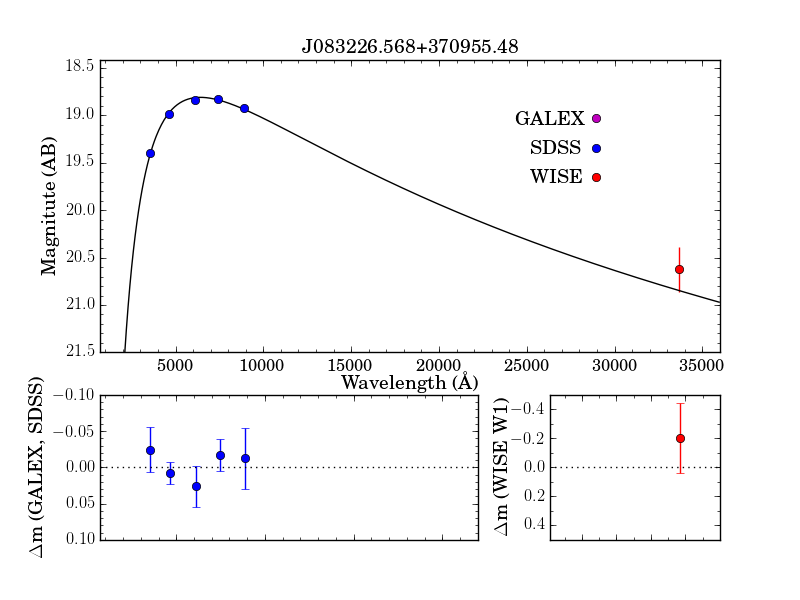}
\includegraphics[width=.49\linewidth,angle=0]{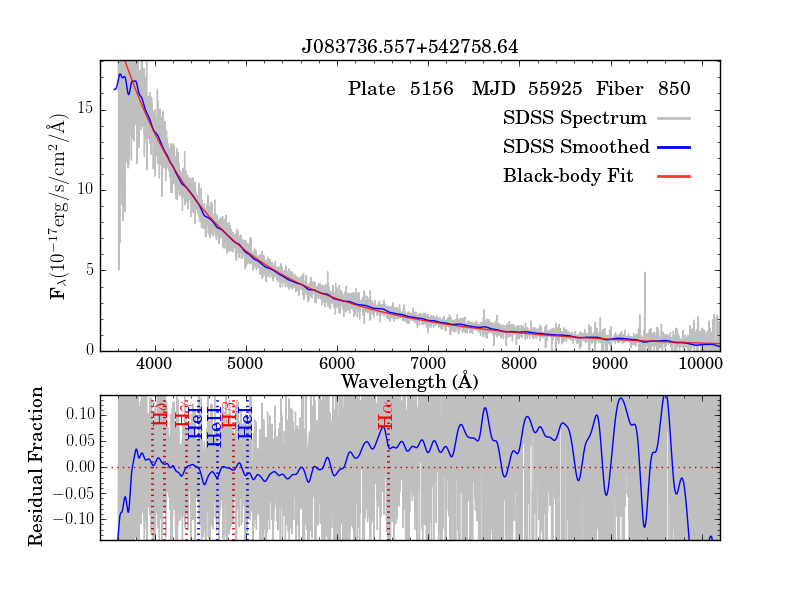}
\includegraphics[width=.49\linewidth,angle=0]{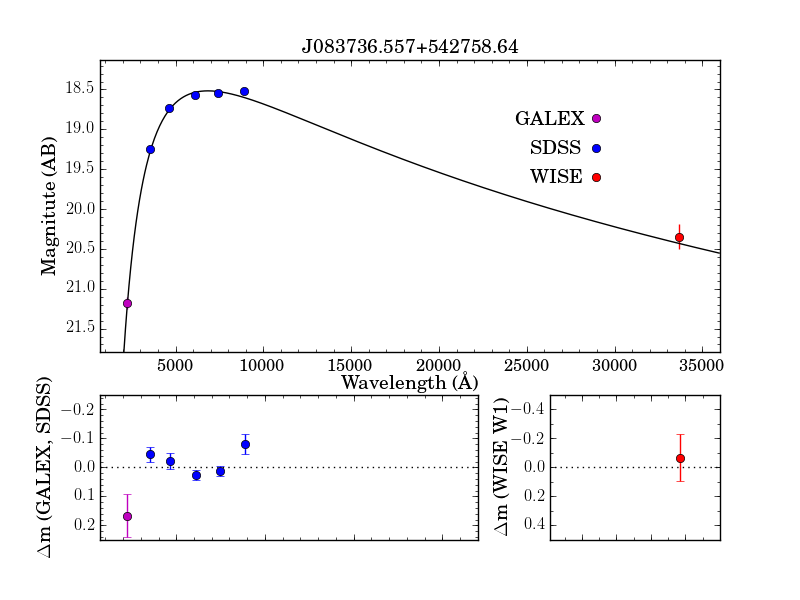}
\end{center}
\end{figure}

\clearpage
\begin{figure}
\begin{center}
\includegraphics[width=.49\linewidth,angle=0]{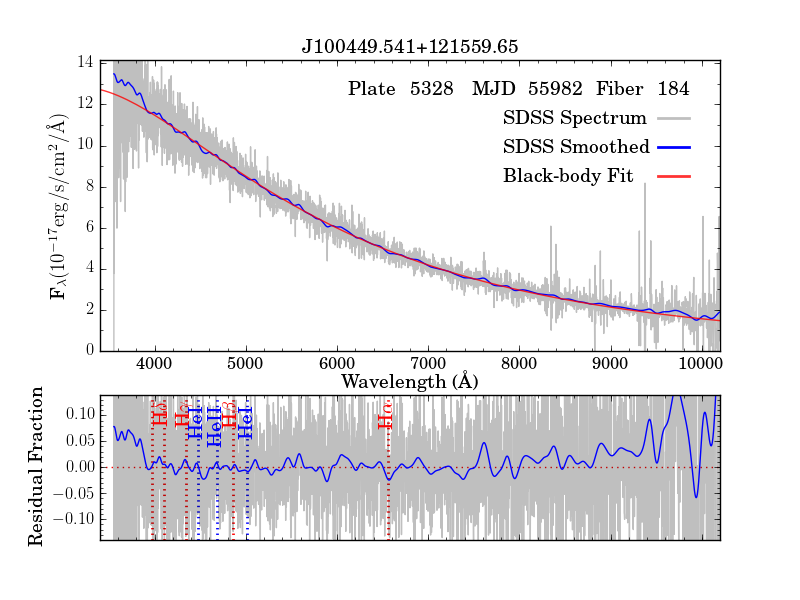}
\includegraphics[width=.49\linewidth,angle=0]{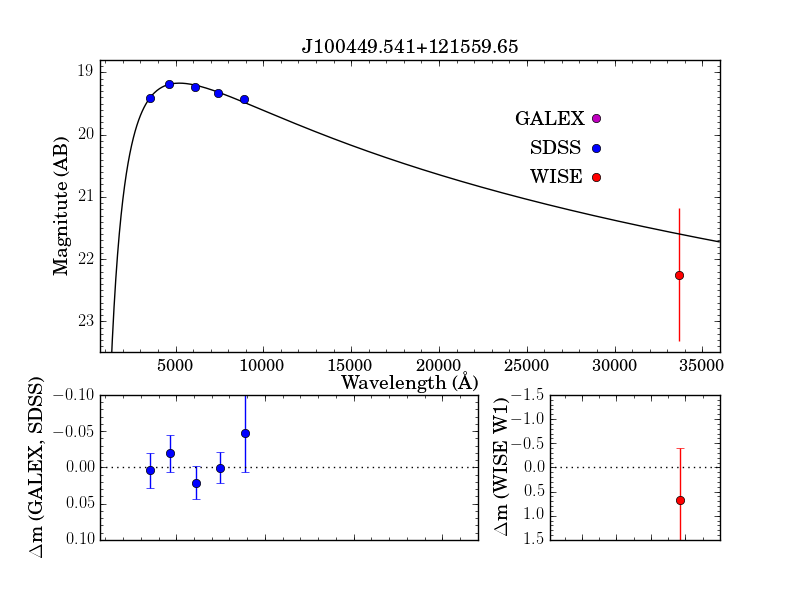}
\includegraphics[width=.49\linewidth,angle=0]{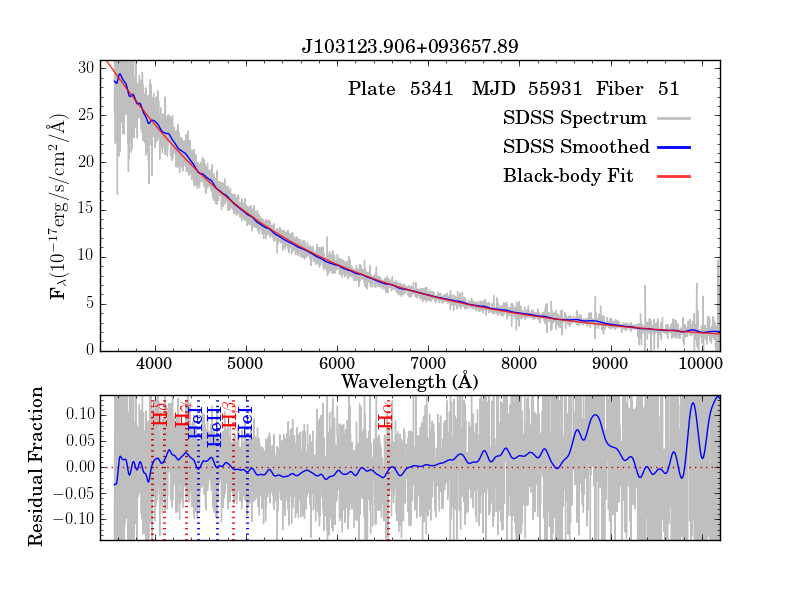}
\includegraphics[width=.49\linewidth,angle=0]{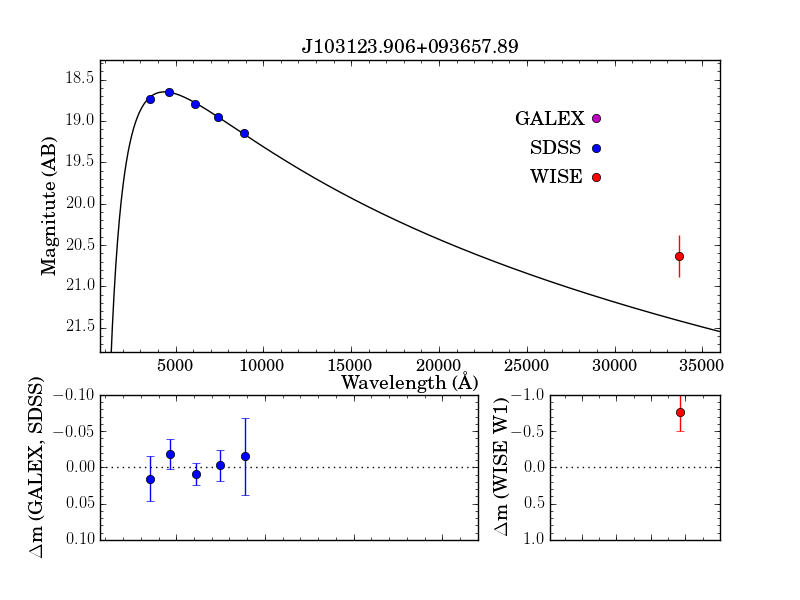}
\includegraphics[width=.49\linewidth,angle=0]{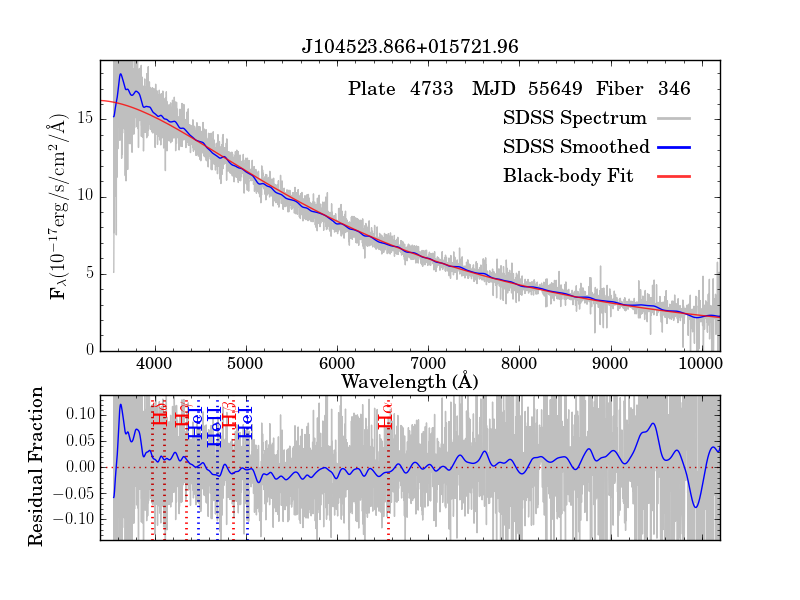}
\includegraphics[width=.49\linewidth,angle=0]{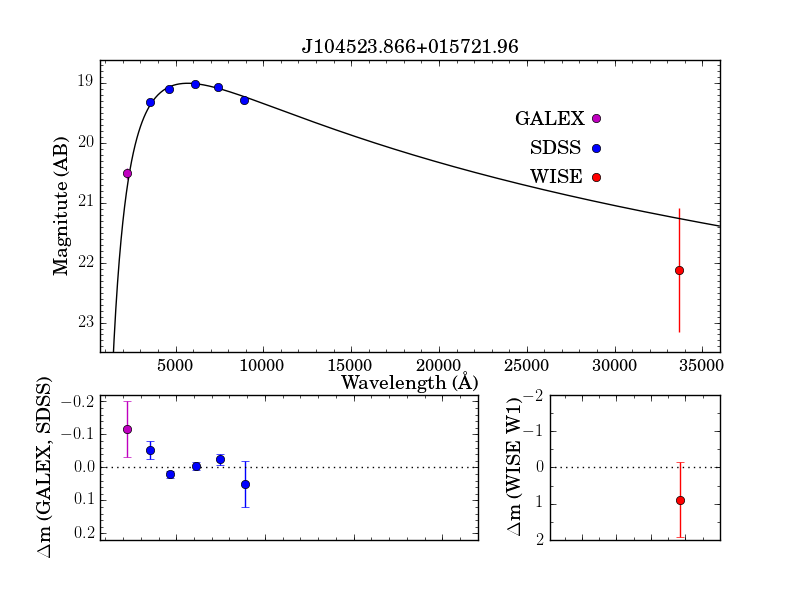}
\includegraphics[width=.49\linewidth,angle=0]{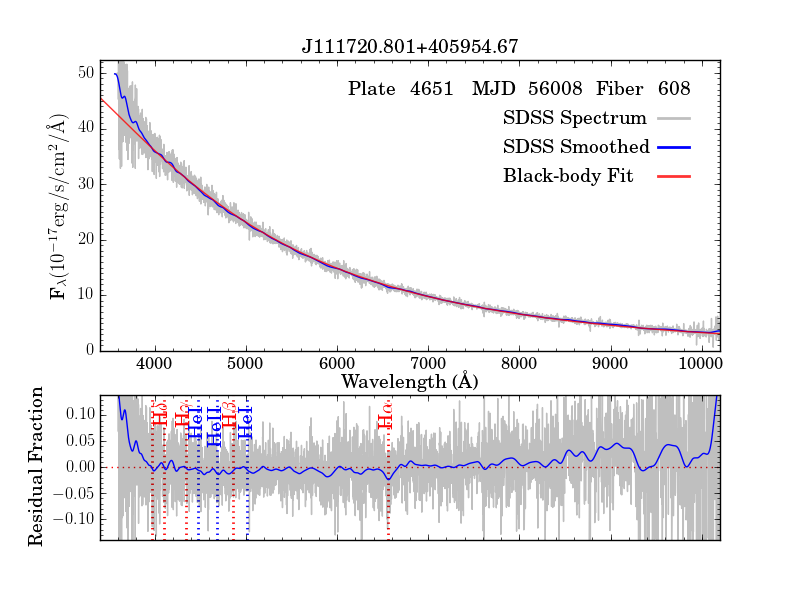}
\includegraphics[width=.49\linewidth,angle=0]{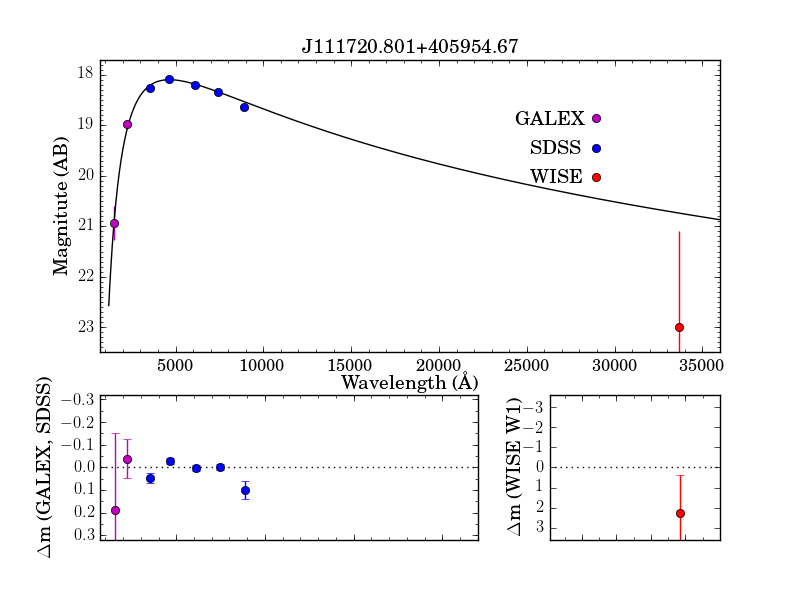}
\end{center}
\end{figure}
\clearpage

\begin{figure}
\begin{center}
\includegraphics[width=.49\linewidth,angle=0]{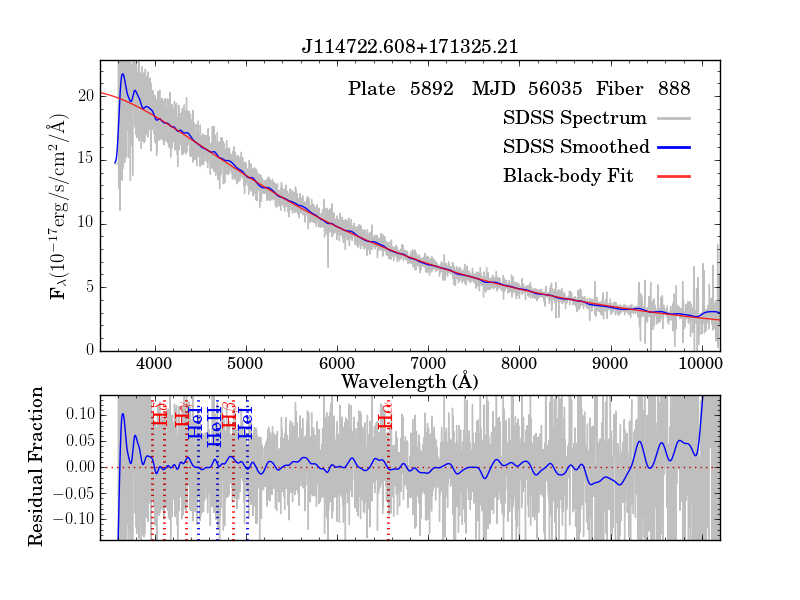}
\includegraphics[width=.49\linewidth,angle=0]{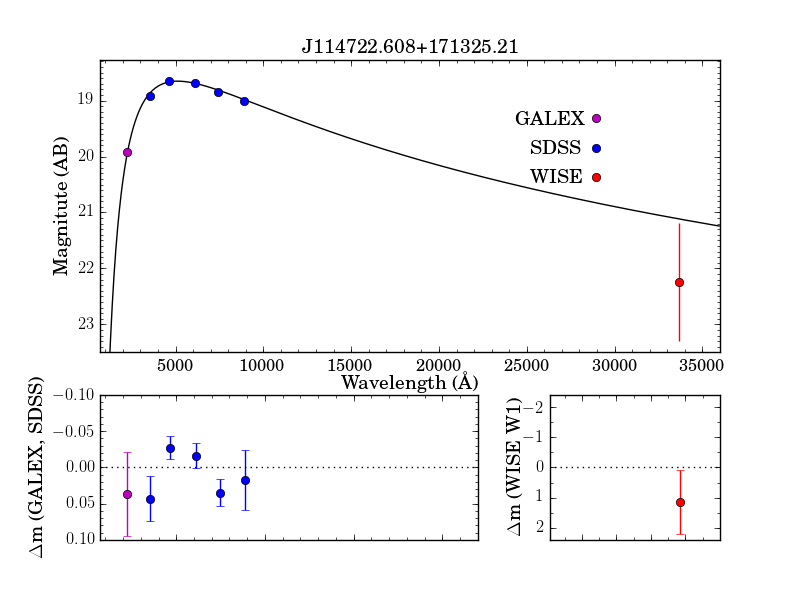}
\includegraphics[width=.49\linewidth,angle=0]{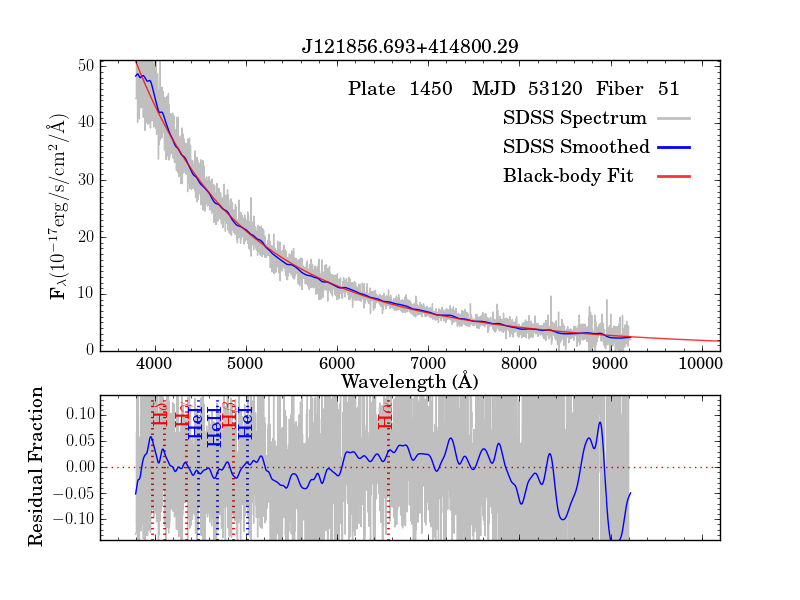}
\includegraphics[width=.49\linewidth,angle=0]{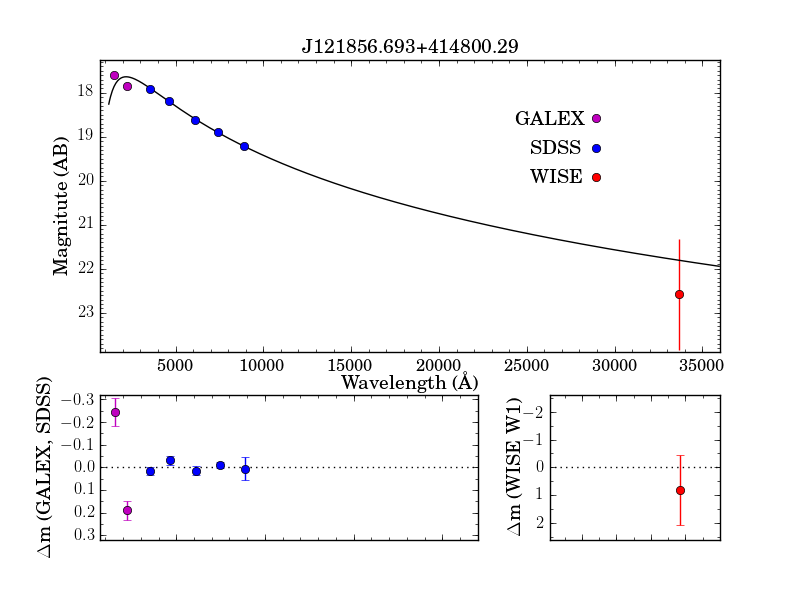}
\includegraphics[width=.49\linewidth,angle=0]{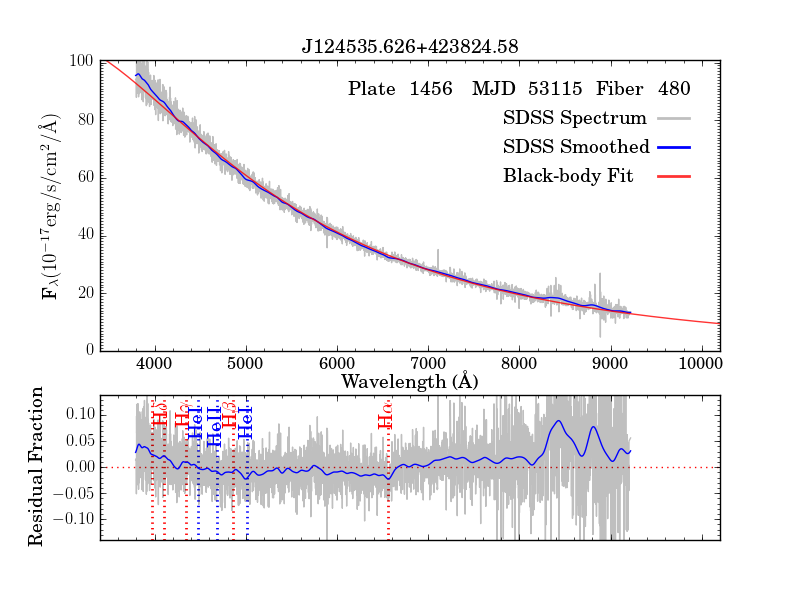}
\includegraphics[width=.49\linewidth,angle=0]{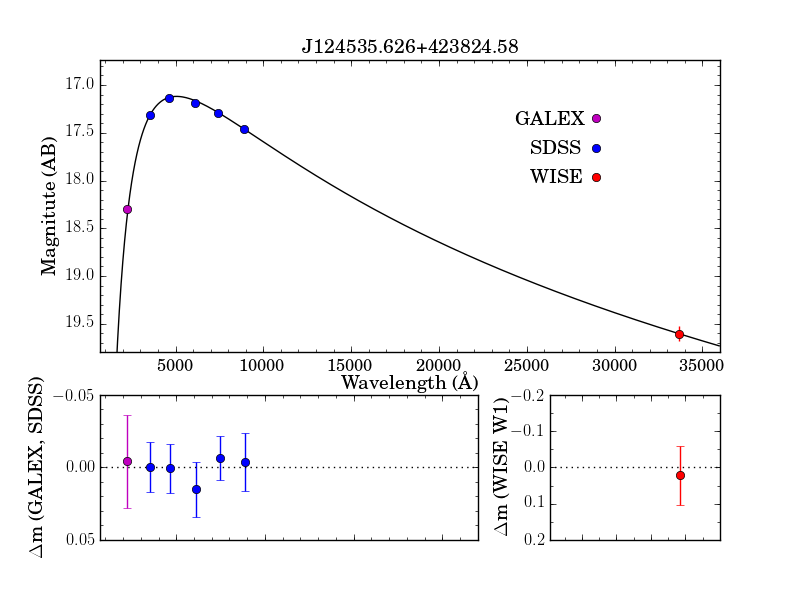}
\includegraphics[width=.49\linewidth,angle=0]{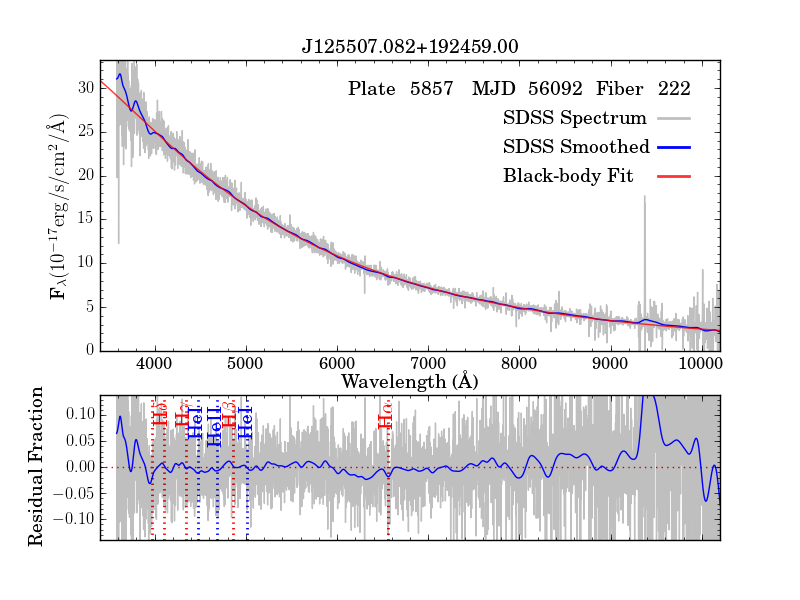}
\includegraphics[width=.49\linewidth,angle=0]{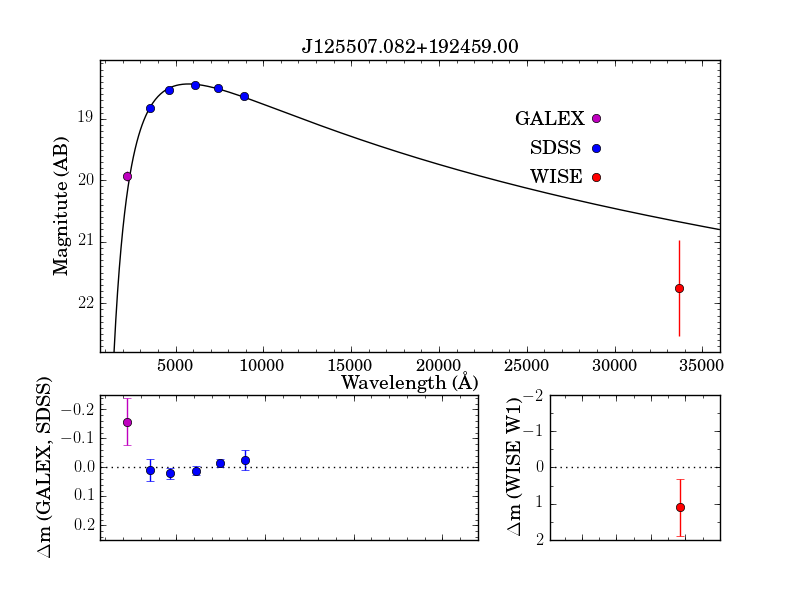}
\end{center}
\end{figure}

\clearpage
\begin{figure}
\begin{center}
\includegraphics[width=.49\linewidth,angle=0]{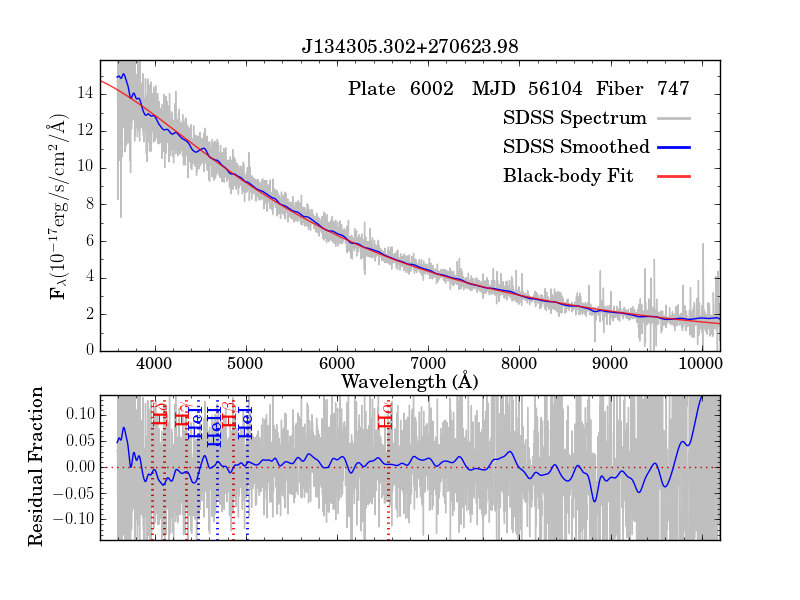}
\includegraphics[width=.49\linewidth,angle=0]{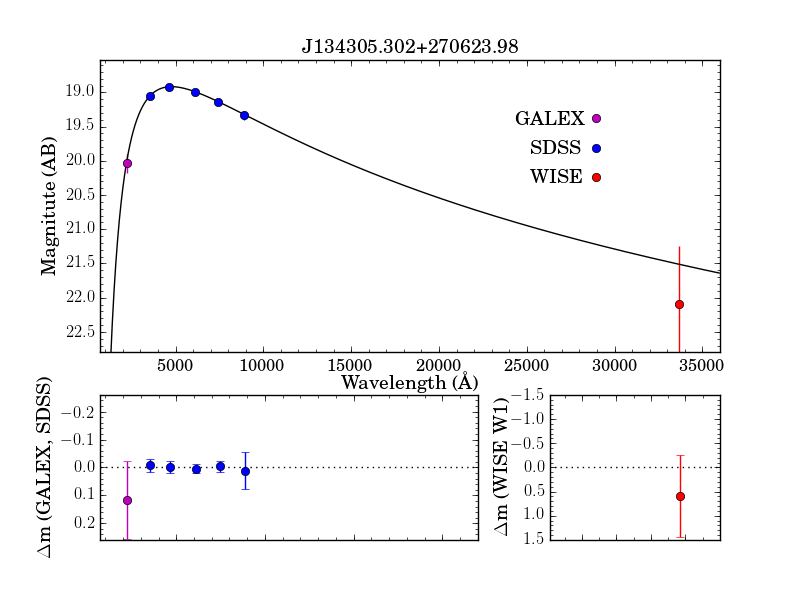}
\includegraphics[width=.49\linewidth,angle=0]{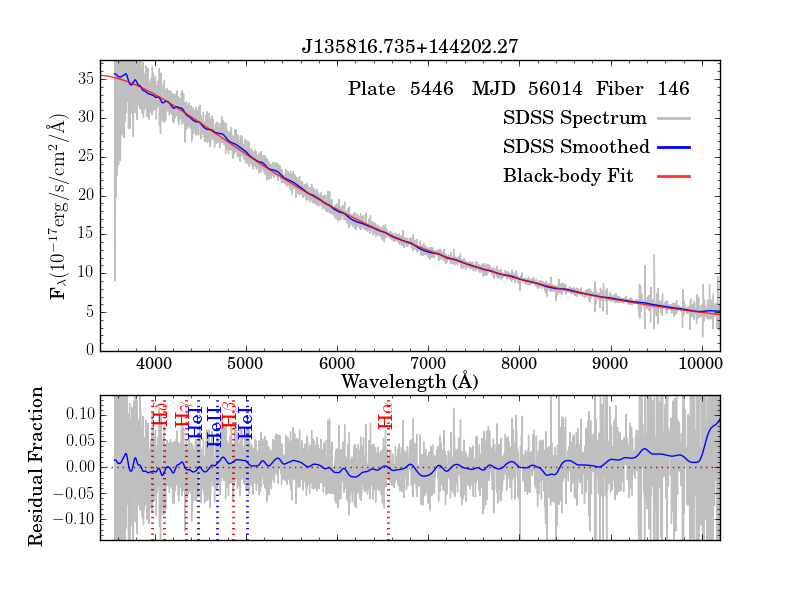}
\includegraphics[width=.49\linewidth,angle=0]{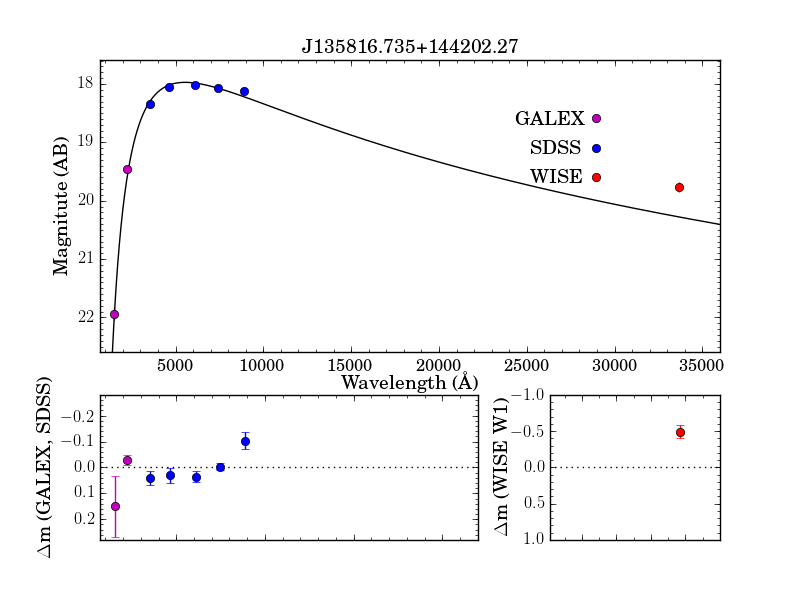}
\includegraphics[width=.49\linewidth,angle=0]{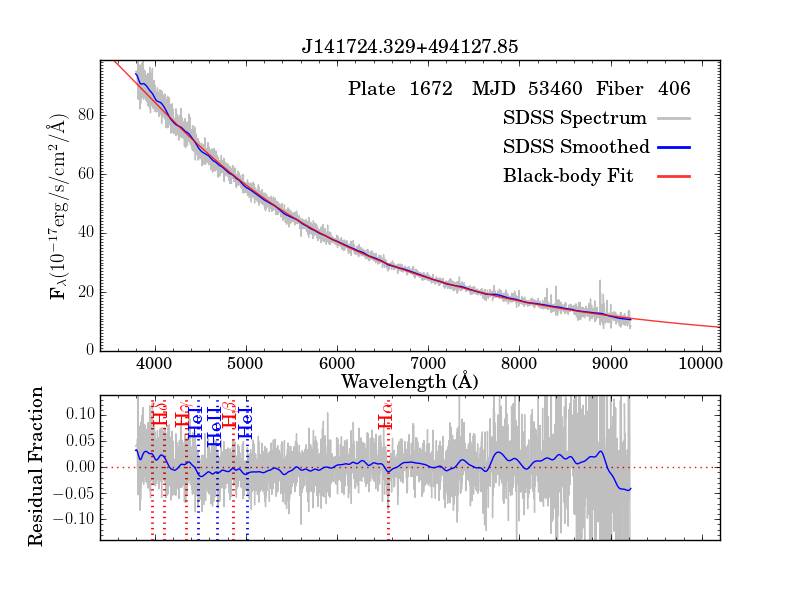}
\includegraphics[width=.49\linewidth,angle=0]{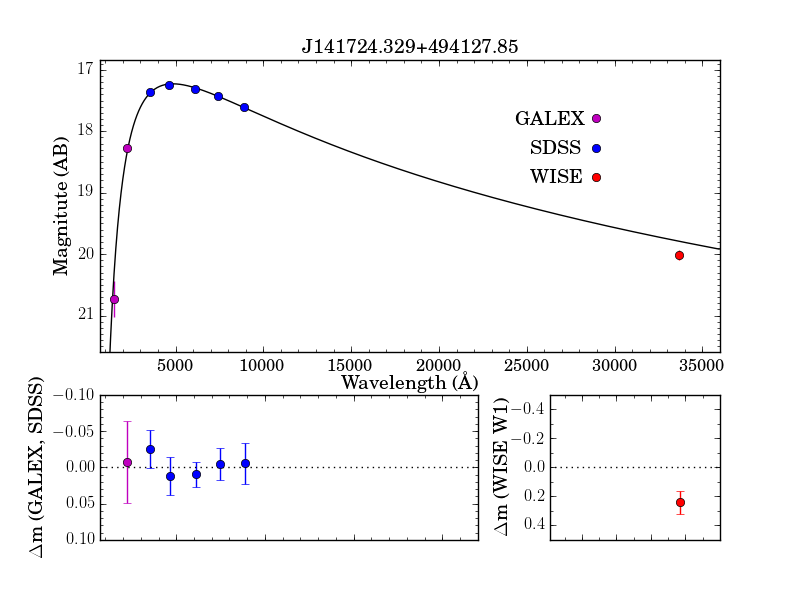}
\includegraphics[width=.49\linewidth,angle=0]{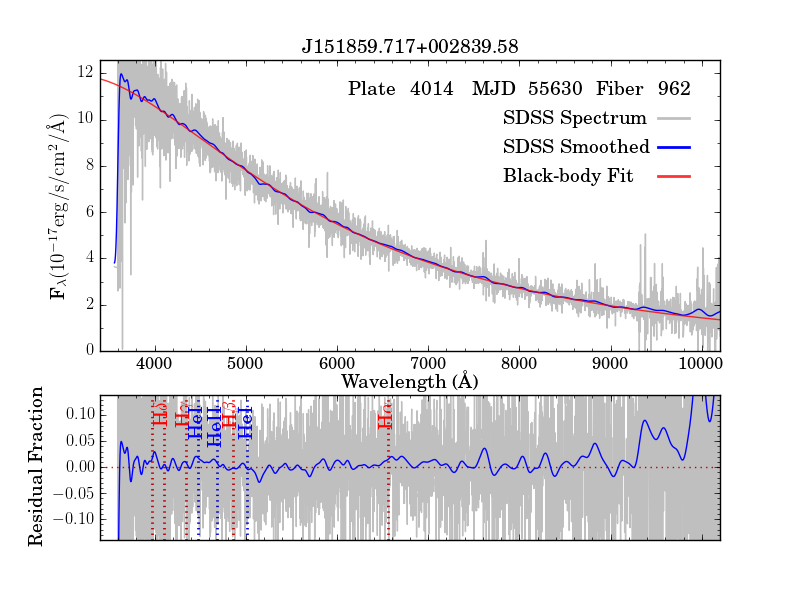}
\includegraphics[width=.49\linewidth,angle=0]{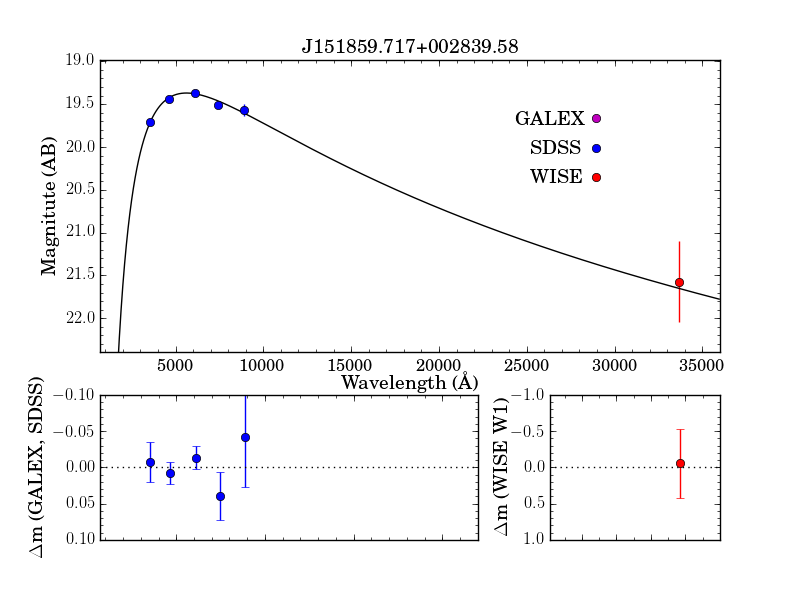}
\end{center}
\end{figure}

\clearpage
\begin{figure}
\begin{center}
\includegraphics[width=.49\linewidth,angle=0]{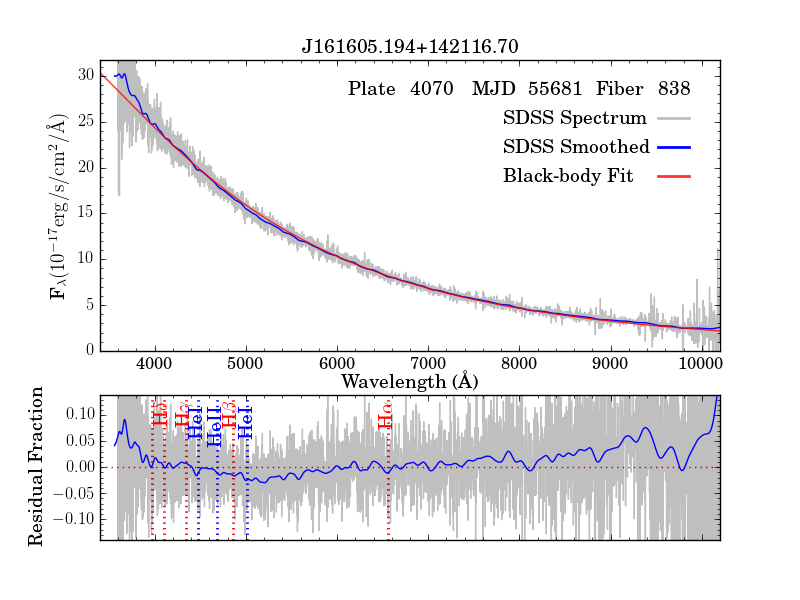}
\includegraphics[width=.49\linewidth,angle=0]{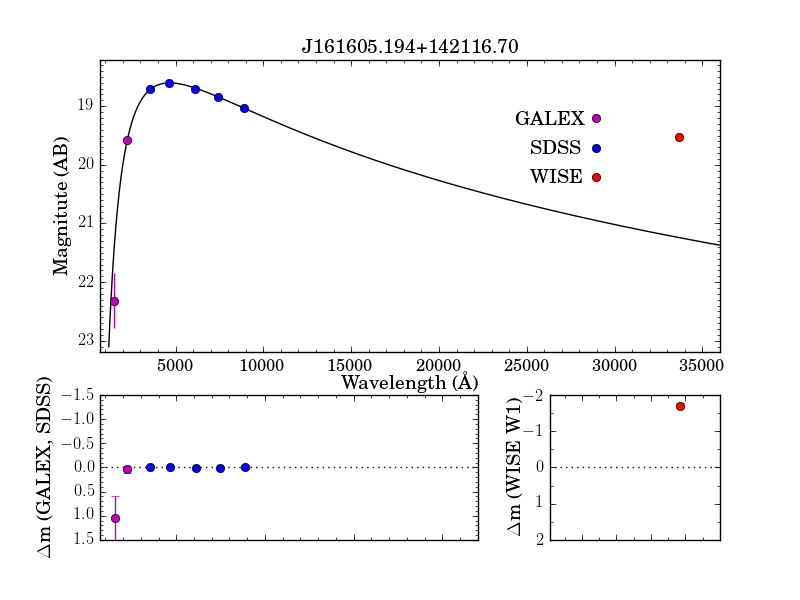}
\includegraphics[width=.49\linewidth,angle=0]{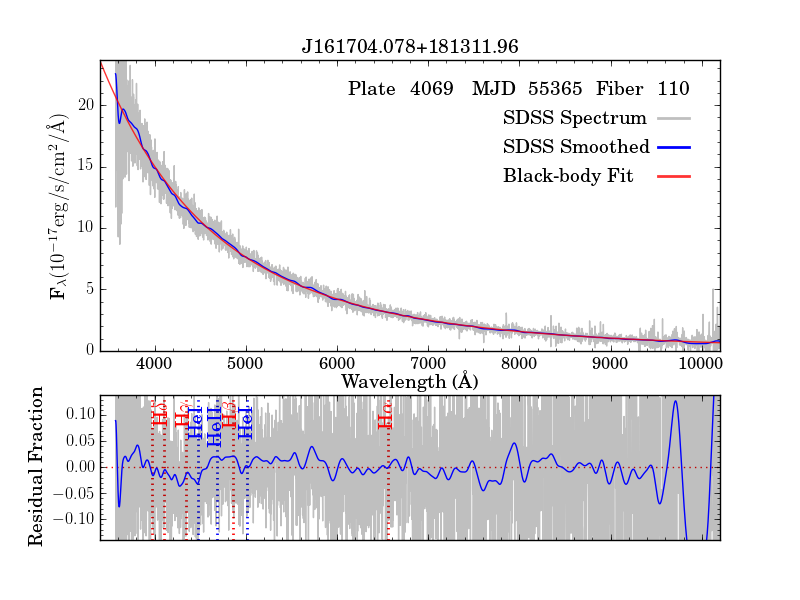}
\includegraphics[width=.49\linewidth,angle=0]{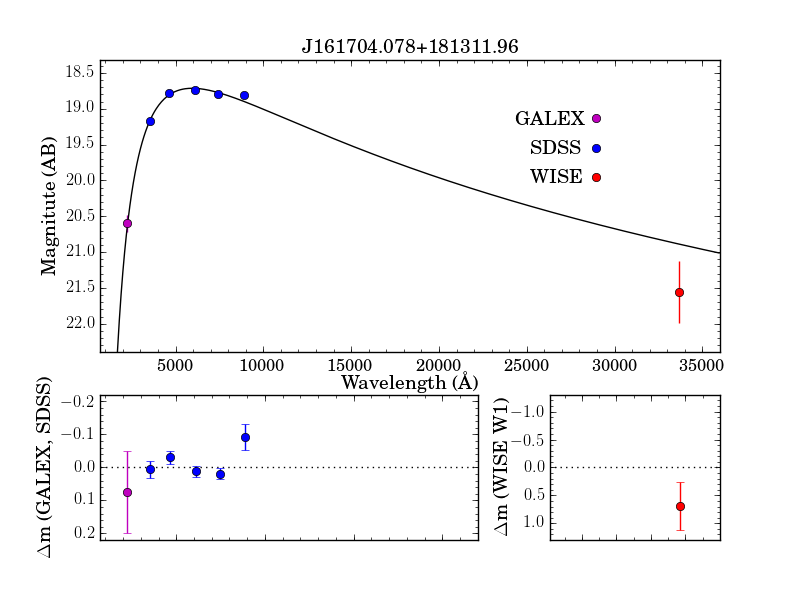}
\includegraphics[width=.49\linewidth,angle=0]{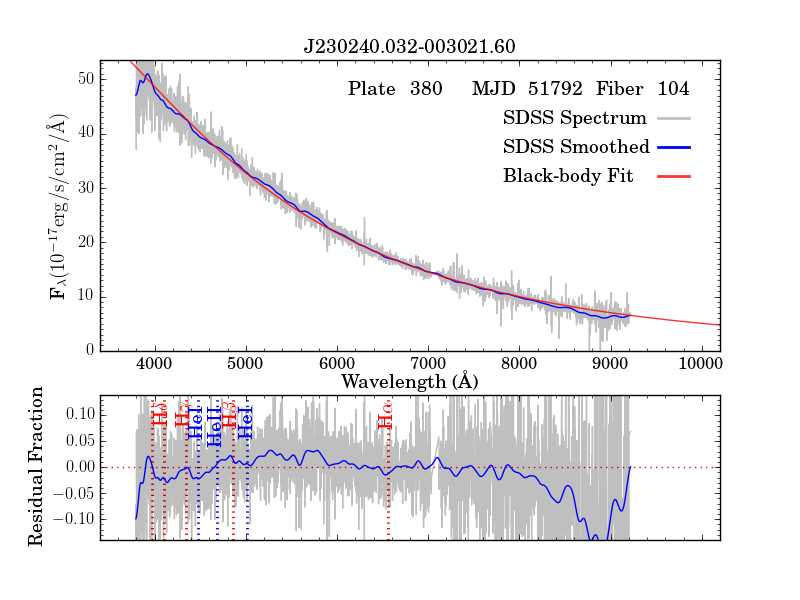}
\includegraphics[width=.49\linewidth,angle=0]{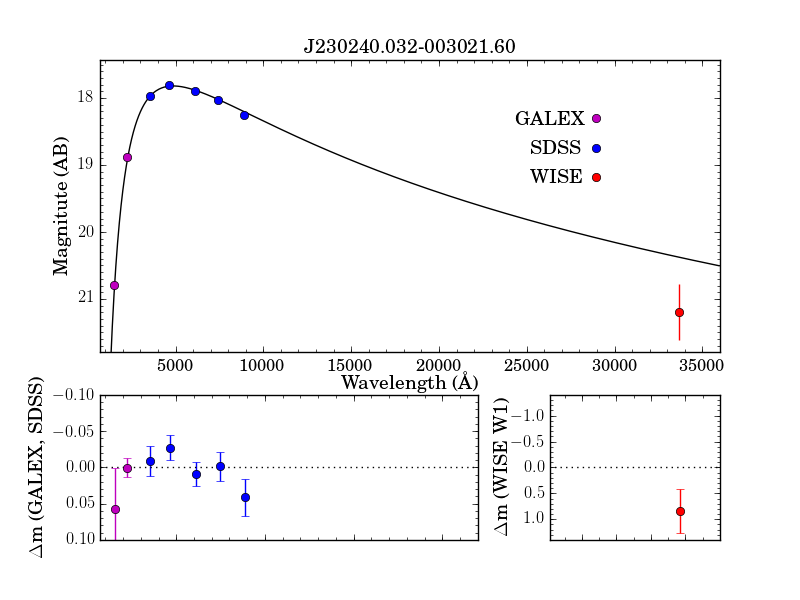}
\end{center}
\end{figure}

\clearpage
\begin{figure}
\includegraphics[width=.7\linewidth,angle=0]{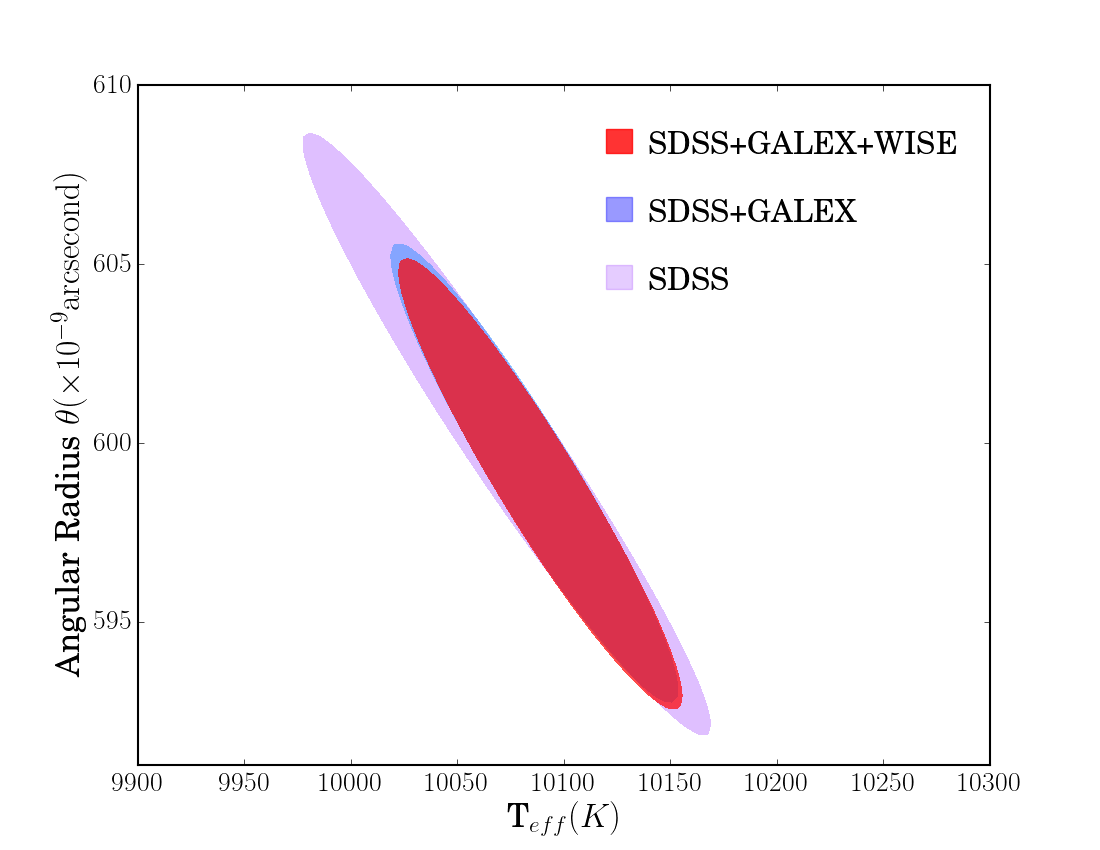}
\caption{
Examples of the black-body fit, the fit for J124535.626$+$423824.58.
One sigma outer contour (thin coloured)
is the fit to SDSS five-colour photometric data alone and
inner contour (thicker coloured) is the fit GALEX UV data included.
Inclusion of WISE IR data does not change the contour.
}
\end{figure}

\clearpage
\begin{figure}
\includegraphics[width=.49\linewidth,angle=0]{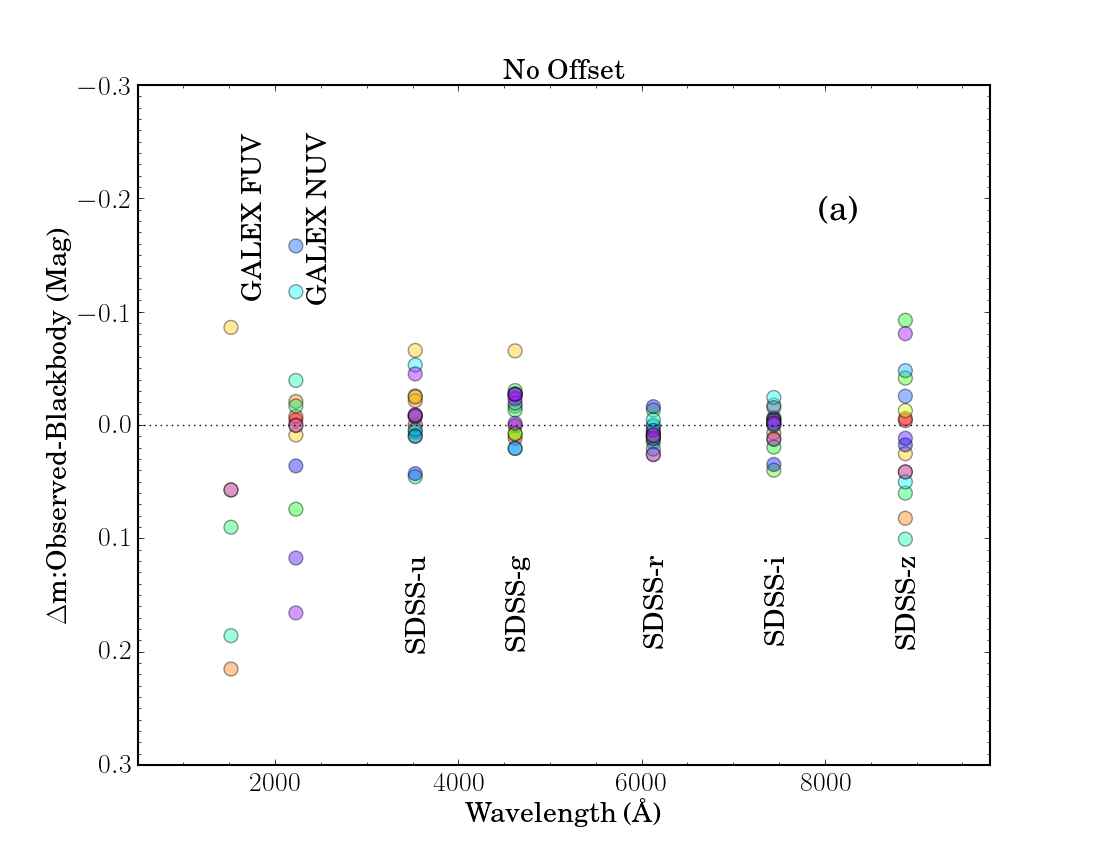}
\includegraphics[width=.49\linewidth,angle=0]{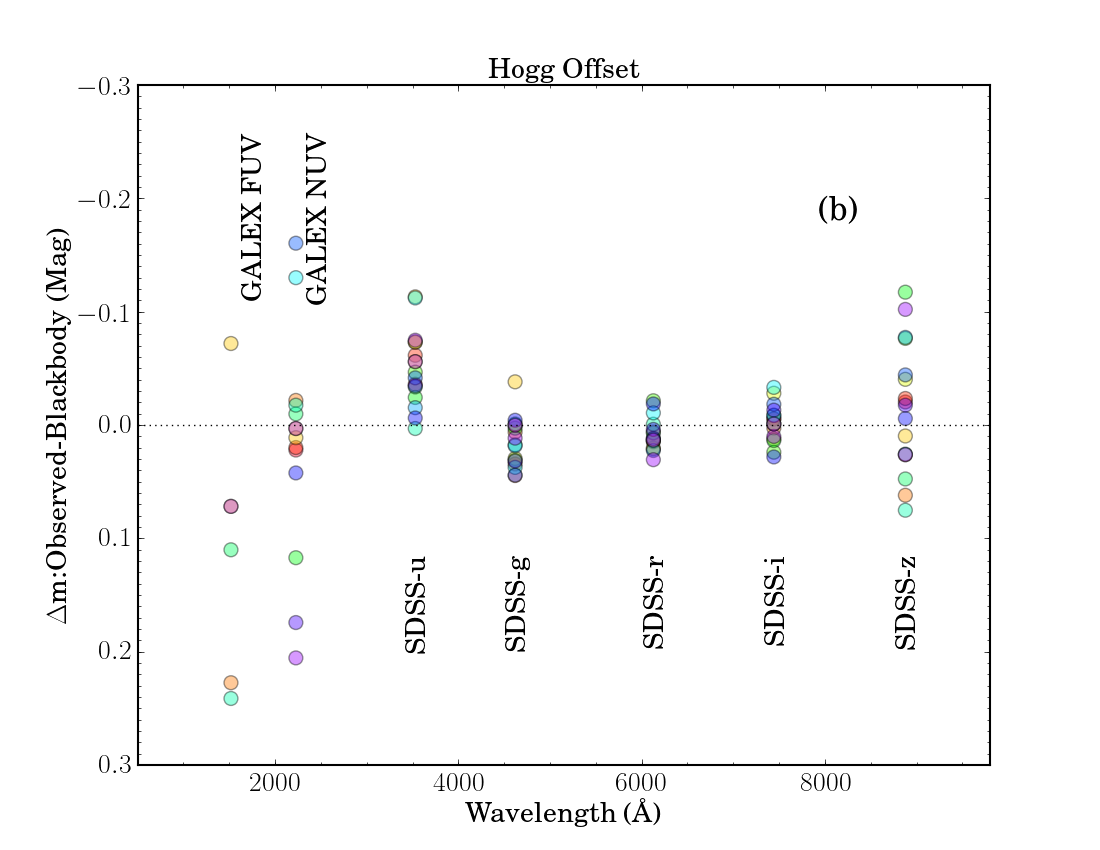}
\includegraphics[width=.49\linewidth,angle=0]{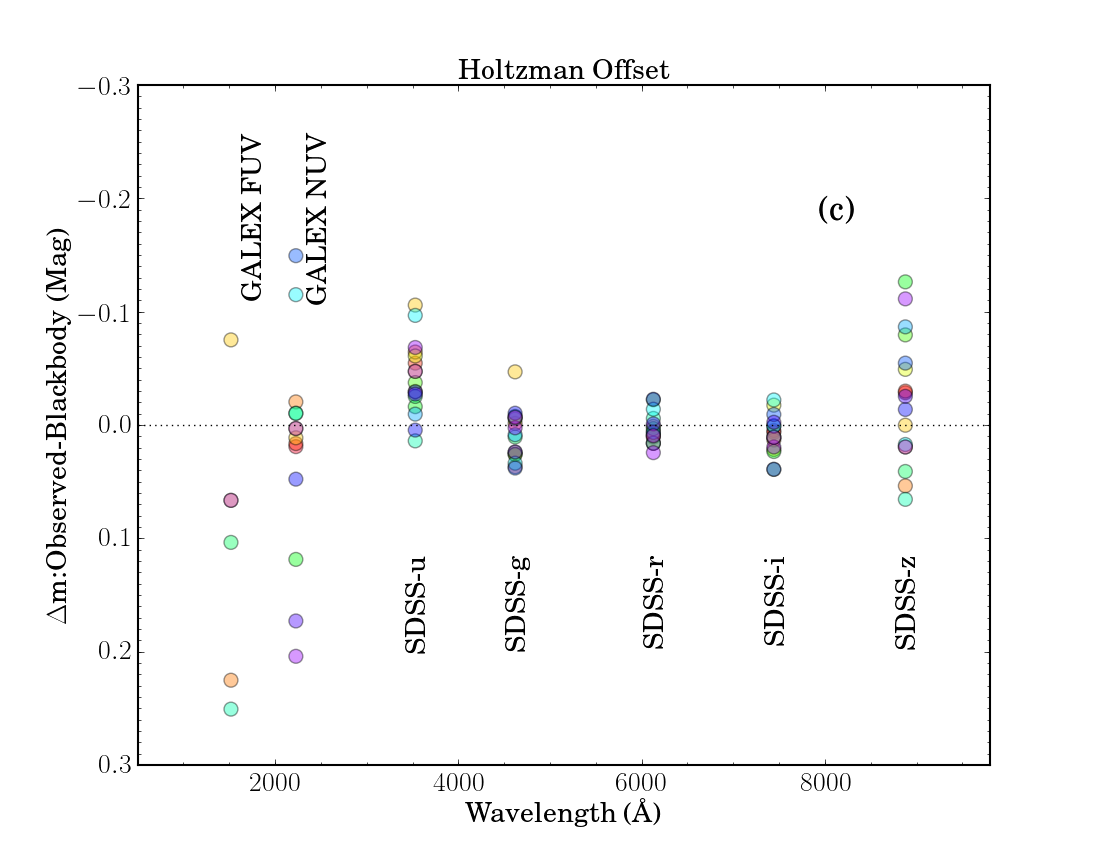}
\caption{
(a) Residuals of the SDSS five-band photometric data and
 the GALEX data from the black body fits in magnitude.
(b) The SDSS photometric data are corrected for the offset proposed by
Hogg and re-fitting is made to give the best fit. The figure shows residuals
in magnitude.
(c) The offset is applied to SDSS photometry as suggested by Holtzman et al. (2009)
based on the CALSPEC zero point. Another re-fitting is made to give the best
black-body fits. The figure shows residuals
in magnitude.
}
\end{figure}
